\documentclass{article}
\usepackage[legalpaper, margin=1in]{geometry}
\usepackage{tabularx}
\usepackage{graphicx} 
\usepackage{authblk}
\usepackage{hyperref} 
\usepackage{wrapfig}
\usepackage{adjustbox}
\usepackage{indentfirst}
\usepackage{anyfontsize}
\usepackage{paracol}
\globalcounter{figure}
\globalcounter{table}
\usepackage{anyfontsize}
\usepackage{subcaption}
\usepackage{subcaption}

\usepackage{fancyhdr}

\usepackage[labelsep=period]{caption}
\begin{document}

\pagestyle{fancy}
\fancyhf{}
\fancyhead[L]{\textit{Material Girl} Launch Report}
\fancyhead[R]{}
\fancyfoot[R]{\parbox[c]{90px}{%
    \includegraphics[width=90px]{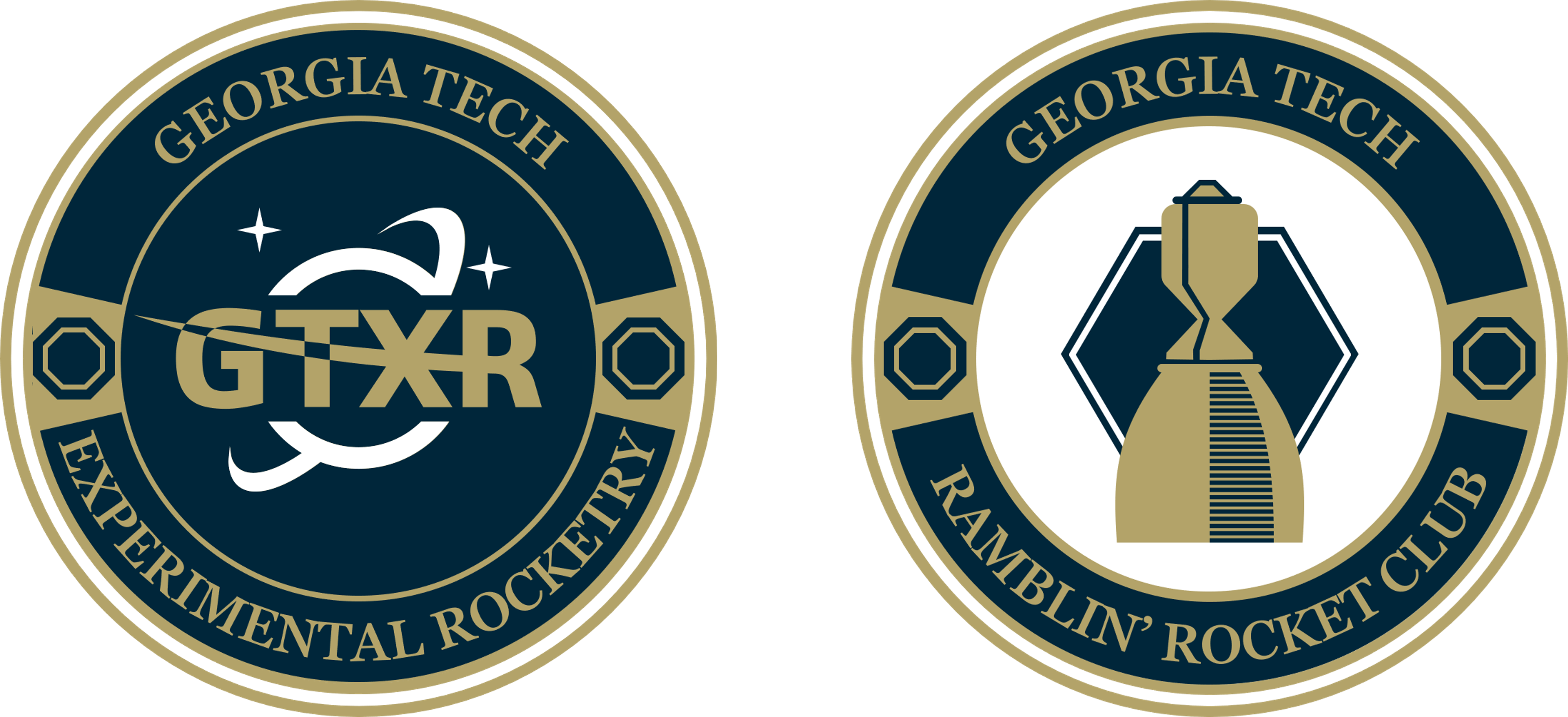}
    }
}%

\fancyfoot[L]{
\parbox[c]{90px}{%
    \thepage
    }
}

\begin{titlepage}

\centering
\vfill
{\fontsize{52}{58}\selectfont{\textit{Material Girl}\\[16pt] Launch Report}}
\vfill
\includegraphics[height=4in]{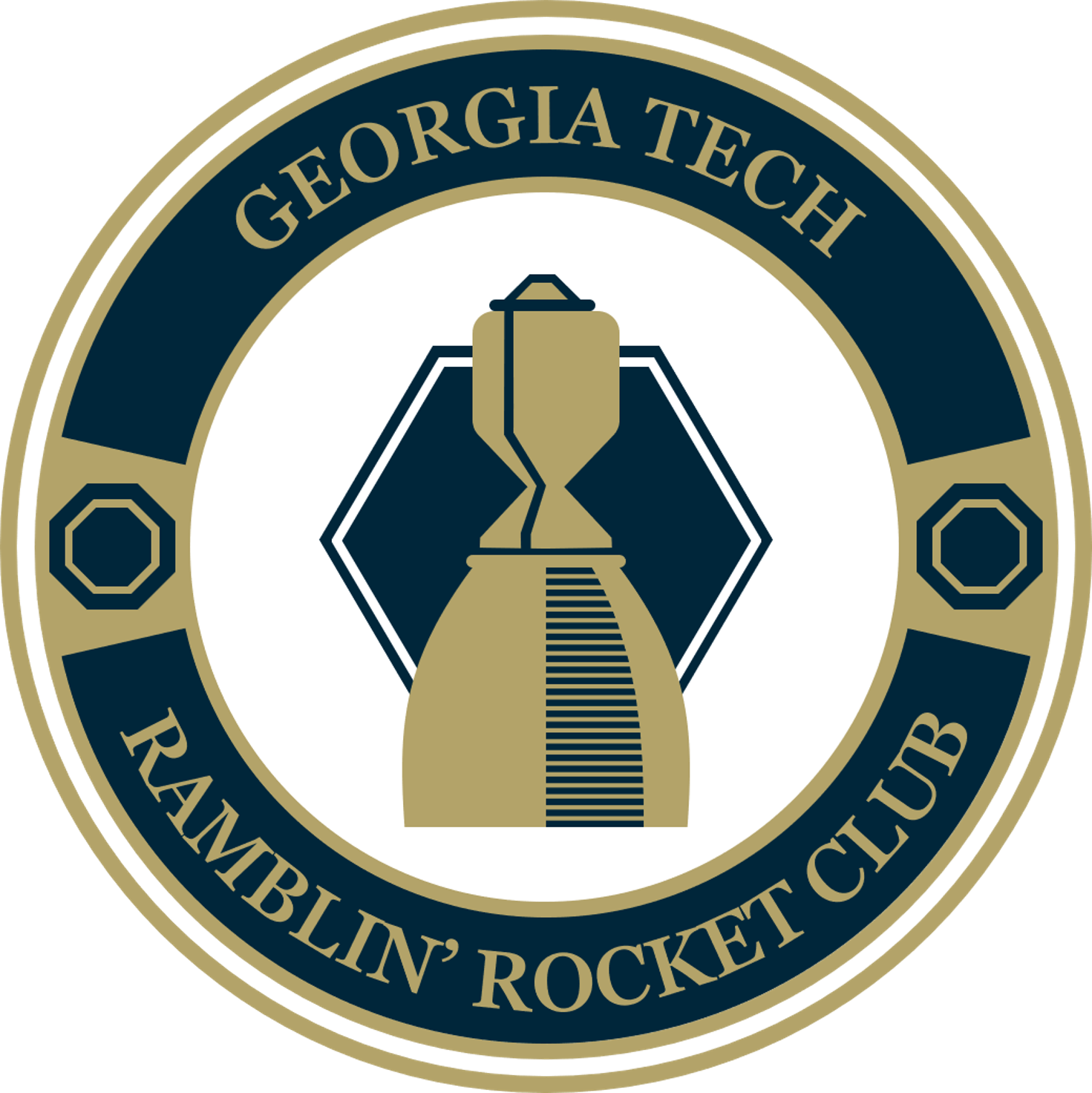}\\
\vspace{0.75in}
\includegraphics[height=4in]{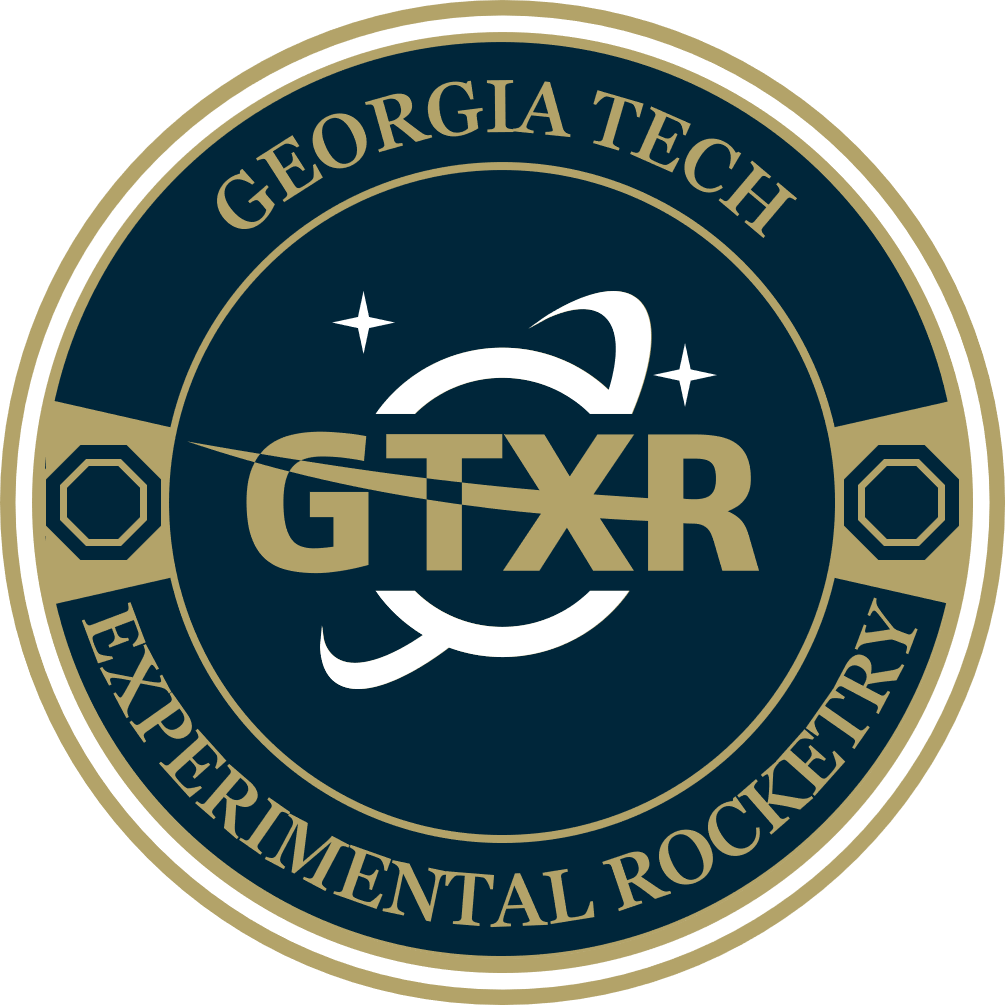}

\vfill
{\huge Ramblin' Rocket Club \\ Georgia Institute of Technology\\}
\vfill
{Parth Garud, Connor Johnson, and Alfonso Lagares de Toledo}
\end{titlepage}

\tableofcontents

\newpage

\section{Introduction}
On July 8th, 2023, Georgia Tech Experimental Rocketry (GTXR), launched a two-stage amateur rocket simulated to an altitude of 220,000 feet above ground level (AGL). The launch took place at Friends of Amateur Rocketry (FAR) in Randsburg, California. The two-stage sounding rocket, named \textit{Material Girl}, was a vehicle designed and built over the 2022-2023 academic year by GTXR. \textit{Material Girl} was an optimized, redesigned, and improved version of the similarly-sized \textit{Mr. Blue Sky} vehicle flown by GTXR on July 9th, 2022.\\

\subsection{Team Background}
Georgia Tech Experimental Rocketry is a team within the Ramblin' Rocket Club, a registered student organization at the Georgia Institute of Technology. The group was founded in 2018 with the goal of sending a two-stage rocket to the Karman Line (100 km AGL), a commonly recognized boundary of outer space. The team has designed, manufactured, and launched four two-stage sounding rockets since its inception. The initial vehicle, \textit{Sustain Alive}, was launched at the Spaceport America Cup in 2019. In 2019, the team started the development of \textit{Rubberband Man}, its next two-stage vehicle which completed development and launched during the summer of 2021. The team then developed \textit{Mr. Blue Sky}, iterating on previous designs and technologies and launching the vehicle in the summer of 2022. The lessons learned by the team through these launches culminated in the design, manufacturing, and launch of \textit{Material Girl} during the 2022-2023 school year.

\subsection{Team Technology Development Roadmap}
In order to meet GTXR's goal of sending a two-stage rocket to 100 km AGL, the team has developed sounding rockets with increasing complexity, iteratively improving on previous designs and learning from each launch. \textit{Material Girl} was the fourth-generation vehicle to follow a two-stage architecture, honing the team's expertise in vehicle integration and launch procedures. This vehicle broadly advanced the technology readiness level (TRL) of subsystems across the board. These vehicles have all contributed to the team's goal of ultimately building a vehicle to reach the Karman Line.
\vfill
    \begin{figure}[h!]
        \centering
        \includegraphics[width=\textwidth]{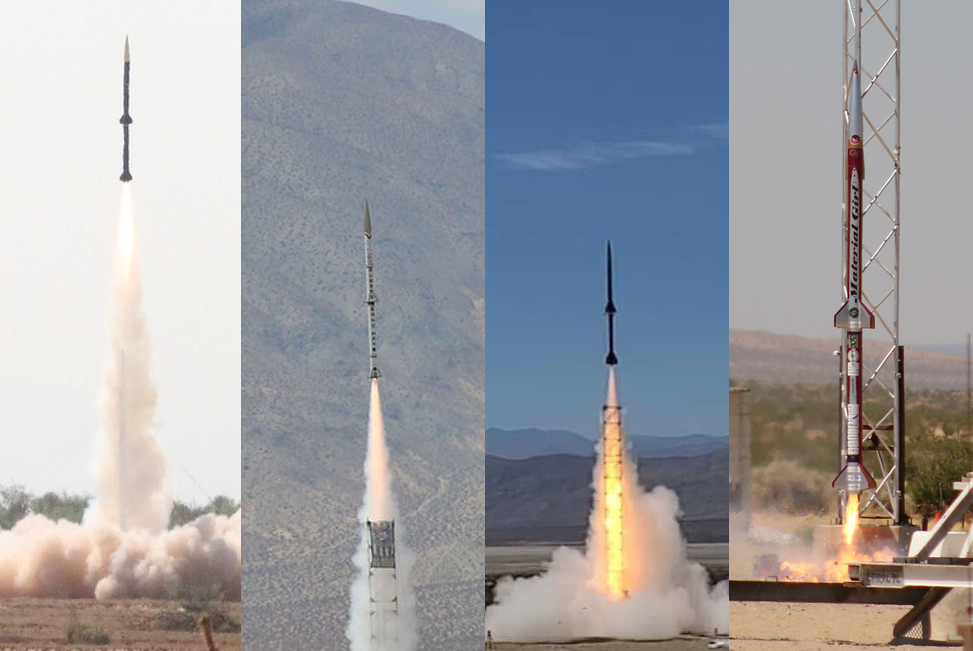}
        \caption{GTXR's four sounding rockets: (L-R) \textit{Sustain Alive} (2019), \textit{Rubberband Man} (2021), \textit{Mr. Blue Sky} (2022), \textit{Material Girl} (2023)}
        \label{fig:mggrouppicture}
    \end{figure}
\vfill
\newpage

\subsection{\textit{Material Girl} Executive Overview}

\columnratio{0.4}
\begin{paracol}{2}

    \vfill
    \begin{figure}[h!]
        \centering
        \includegraphics[height=2.4in,angle=90]{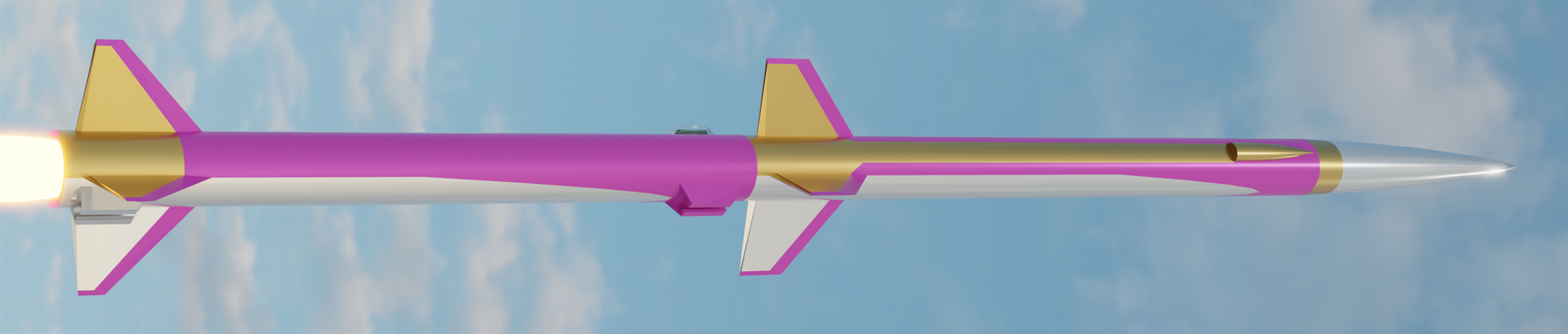}
        \caption{Vehicle rendering.}
        \label{fig:Render}
    \end{figure}
    
    \switchcolumn

    \vfill
    
    \textit{Material Girl} was a two-stage sounding rocket designed and flown during the 2022-2023 academic year. Weighing a total of 197 lbs and spanning 176 inches in length, the vehicle featured many improvements from the team's previous vehicle, \textit{Mr. Blue Sky}. \\
    
    The vehicle featured a Student Researched and Developed (SRAD) P-class motor on each stage, a mechanical staging system, and recovery systems for the booster, sustainer, and nose cone. Flight events were designed to be controlled by a COTS avionics flight computer, with a custom flight computer stored as a payload. The motor grain geometry and fin dimensions were determined through an in-house optimization software derived from RASAero and RocketPy.\\

\vfill

\begin{table}[h!]
    \centering
    \caption{\textit{Material Girl} vehicle dimensions.}
    \begin{tabular}{||cc||} 
     \hline
     \textbf{Parameter} & \textbf{Quantity}\\ [0.5ex] 
     \hline\hline
     Length & 176 in \\ 
     Diameter & 6.17 in \\
     Dry mass & 99 lbm \\
     Propellant mass & 98 lbm \\
     Total mass & 197 lbm \\
     Total CG position & 106 in \\
     Number of fins & 4 \\
     \hline
    \end{tabular}
    \label{table:1}
\end{table}

\vfill

\begin{table}[h!]
\centering
\caption{\textit{Material Girl} simulated performance.}
\begin{tabular}{||c c||} 
 \hline
 \textbf{Parameter} & \textbf{Quantity}\\[0.5ex] 
 \hline\hline
 Expected apogee & 220,000 ft   \\ 
 Maximum speed & Mach 4.1 \\
 Maximum thrust & 2,207 lbf \\
 Total impulse & 20,600 lbf*sec \\
 Burn duration & 14 sec \\
 \hline
\end{tabular}
\label{table:1}
\end{table}

\vfill

\begin{figure}[h!]
        \centering
        \includegraphics[height=2.4in]{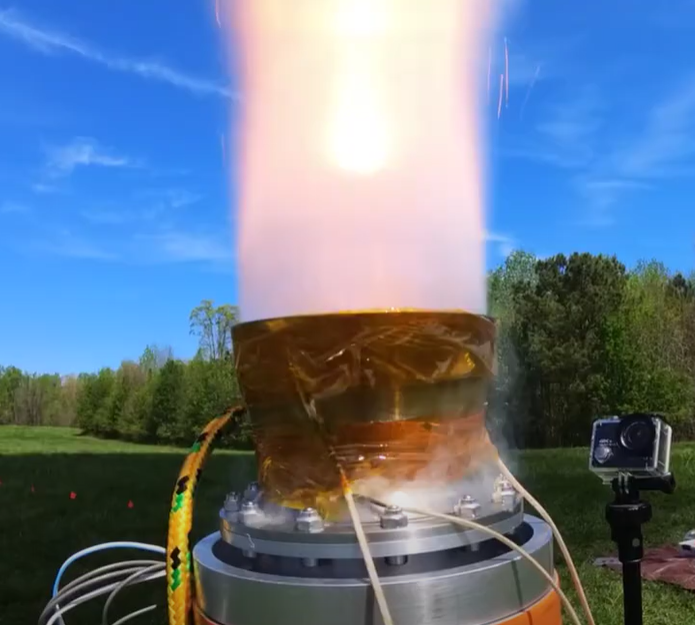}
        \caption{Motor Hot-Fire Test.}
        \label{fig:Motor Hot-Fire Test}
    \end{figure}

\end{paracol}
\newpage

\section{Vehicle Design}
\subsection{Simulations}



A nominal flight trajectory was simulated using the six degrees of freedom (6-DOF) Python-based flight simulation library called RocketPy. RocketPy is open-source, but was heavily modified to provide the capabilities described in this document. This flight was simulated with the expected weight and drag values based on manufactured components, as well as a constant ten miles per hour wind speed. Simulations with full location-based wind profiles were conducted successfully; however, the constant ten miles per hour wind setting was a worst-case scenario condition that was selected to demonstrate the safety of the flight.\\

\begin{figure}[h!]
    \centering
    \includegraphics[width=0.9\textwidth]{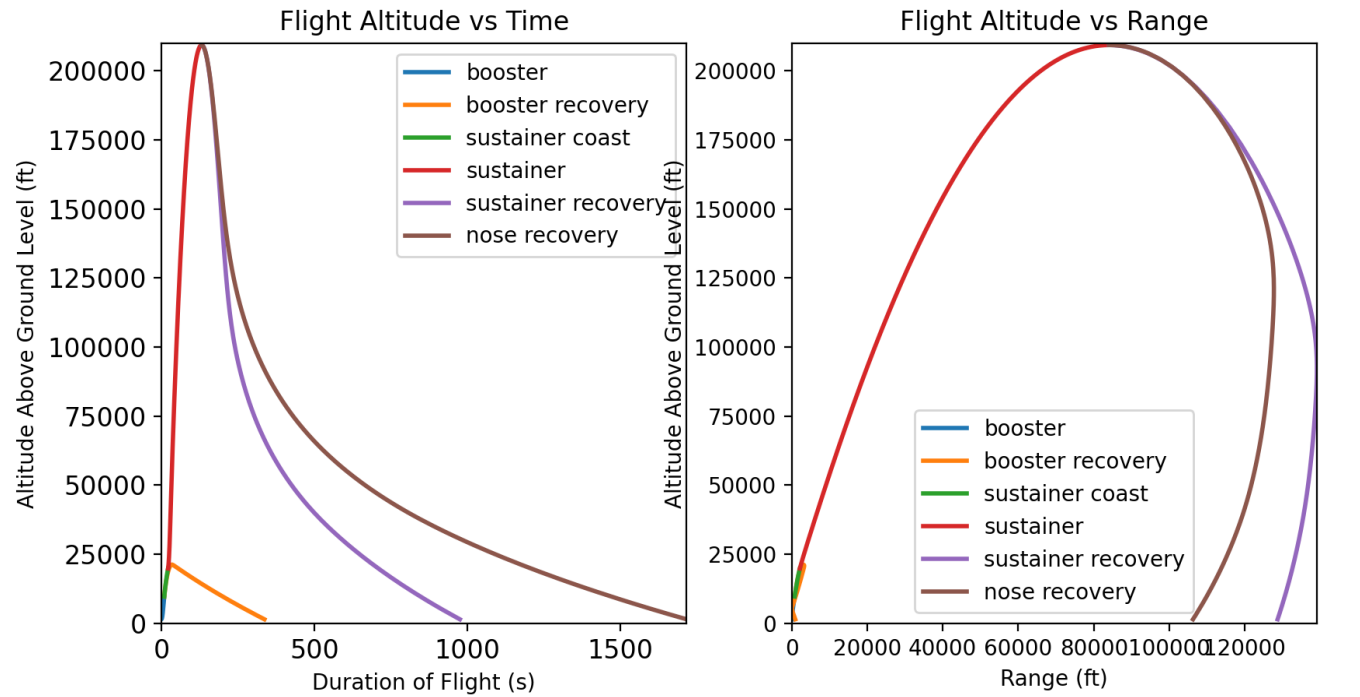}
    \caption{Altitude plot for nominal flight simulation (10 mph of constant wind). Altitude vs Time (left), Altitude vs Range (right). Distance measurements in feet.}
    \label{fig:Altitude}
\end{figure}

\begin{figure}[h!]
    \centering
    \includegraphics[width=0.9\textwidth]{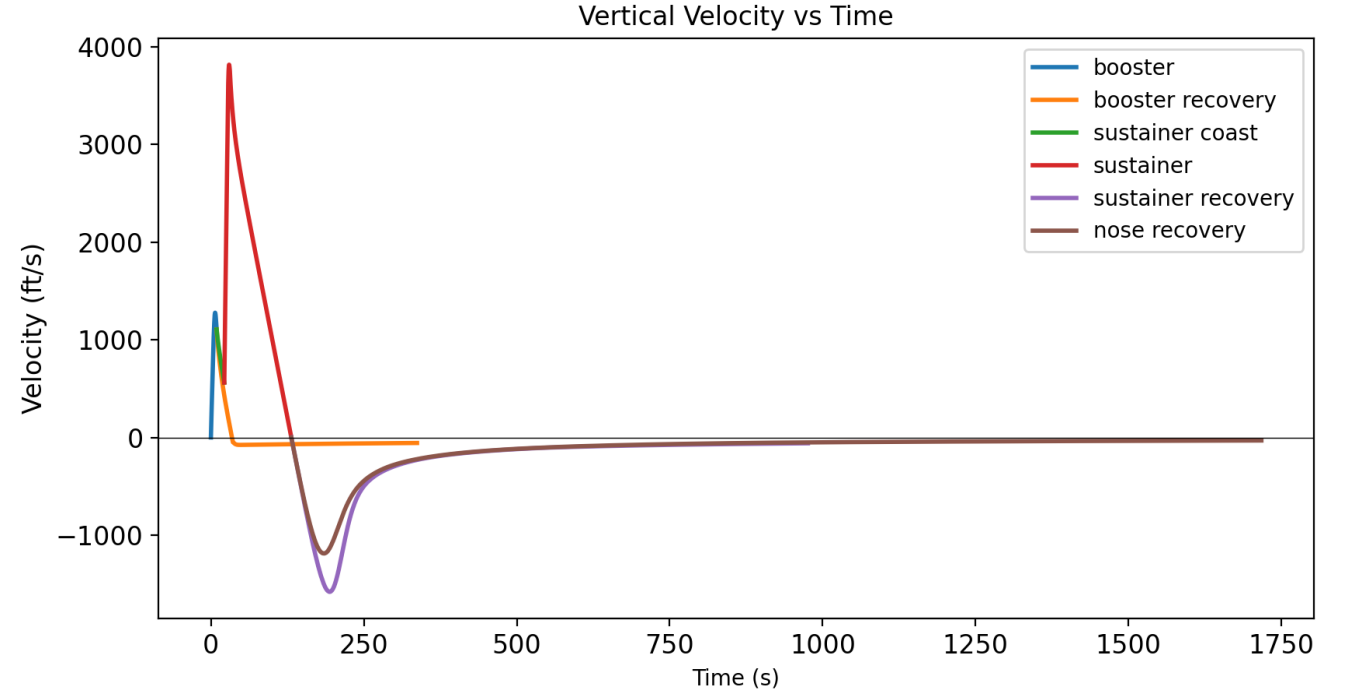}
    \caption{Velocity plot from nominal flight simulation (10 mph constant wind). Velocity is measured in feet per second.}
    \label{fig:Velocity}
\end{figure}

All RocketPy flight simulations are broken up into six events: booster, booster recovery, sustainer coast, sustainer, sustainer recovery, and nose recovery. The booster flight event simulates the ascent of the full rocket until staging. A delay exists between staging and sustainer ignition to give the vehicle’s fins ample time to straighten its flight path and increase the apogee of the vehicle. The sustainer coast event simulates this delay of the unignited sustainer. The sustainer simulation event simulates the ascent of the sustainer section of the rocket during and after the sustainer motor burn. Finally, all recovery simulation events simulate the descent of their respective rocket sections. The state variables of each simulation event are appropriately passed onto the following events. Figure \ref{fig:Altitude} shows relevant altitude data from the nominal flight simulation, including an apogee of approximately 220,000 feet. The color-coded legend shows the six simulation events described earlier. Table \ref{table:NominalFlightEvent} shows a more detailed breakdown of important altitude information, time, and velocity information of nominal flight events. From the same nominal flight simulation, Figure \ref{fig:Velocity} shows relevant vertical velocity information during different flight events. During this simulation, the sustainer reached a peak Mach number of 4.05.
 
\begin{table}[h!]
\centering
\caption{\textit{Material Girl} nominal flight events.}
\begin{tabular}{||c c c c||} 
 \hline
 \textbf{Event} & \textbf{Time (s)} & \textbf{Altitude (m)} & \textbf{Velocity (m/s)} \\ [0.5ex] 
 \hline\hline
 Liftoff & 0 & 0 & 0 \\ 
 Clears Rail &0.42 &18.29 &44.29 \\
 Staging & 9.41 &3128.13 &336.71\\
 Sustainer Ignition &21.91 &6224.27 &171.24\\
 Max Q &28.48 &10608.70 &1117.68\\
 Booster Apogee &34.38 &6655.33 &0\\
 Sustainer Apogee &131.44 &63982.78 &0\\
 Booster Touchdown &336.96 &0 &17.30\\
 Sustainer Touchdown &975.95 &0 &18.30\\
 Nose Cone Touchdown &1717.22 &0 &9.59\\
 \hline
\end{tabular}
\label{table:NominalFlightEvent}
\end{table}

\newpage

Figure \ref{fig:Pitch} below shows pitch angle information for the three ascent simulation events of the nominal flight simulation. Due to the rocket’s stability, \textit{Material Girl} will pitch slightly into the wind during boost. However, this effect is largely unavoidable when ensuring a rocket is adequately stable and safe to fly. Additionally, as seen in Figure \ref{fig:Pitch}, the sustainer only begins to reach a high pitch angle during the end of its ascent. This is due to the decrease in velocity and stability as the rocket reaches its apogee. At this point in the flight, the rocket has slowed down significantly and will deploy parachutes upon reaching
apogee.

\begin{figure}[h!]
    \centering
    \includegraphics[width=0.9\textwidth]{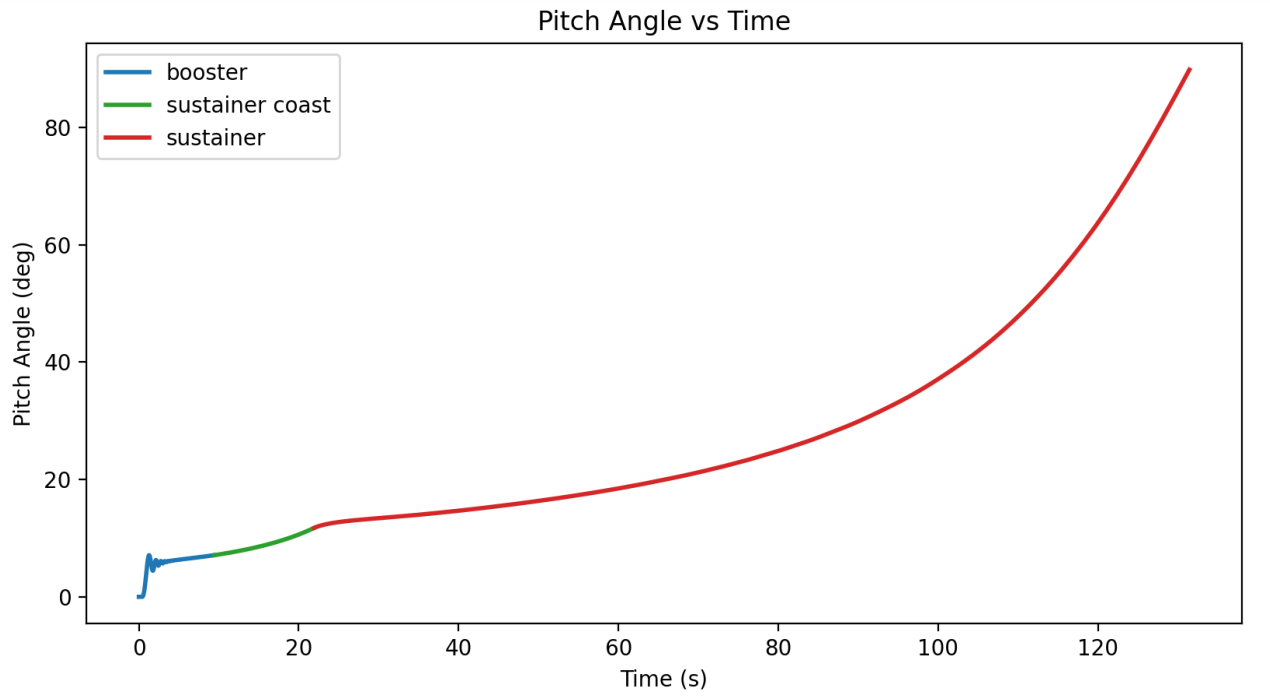}
    \caption{Pitch plot from nominal flight simulation (10 mph constant wind). Pitch is measured in degrees. A zero pitch angle is defined as vertical.}
    \label{fig:Pitch}
\end{figure}

\begin{figure}[h]
    \centering
    \includegraphics[width=0.9\textwidth]{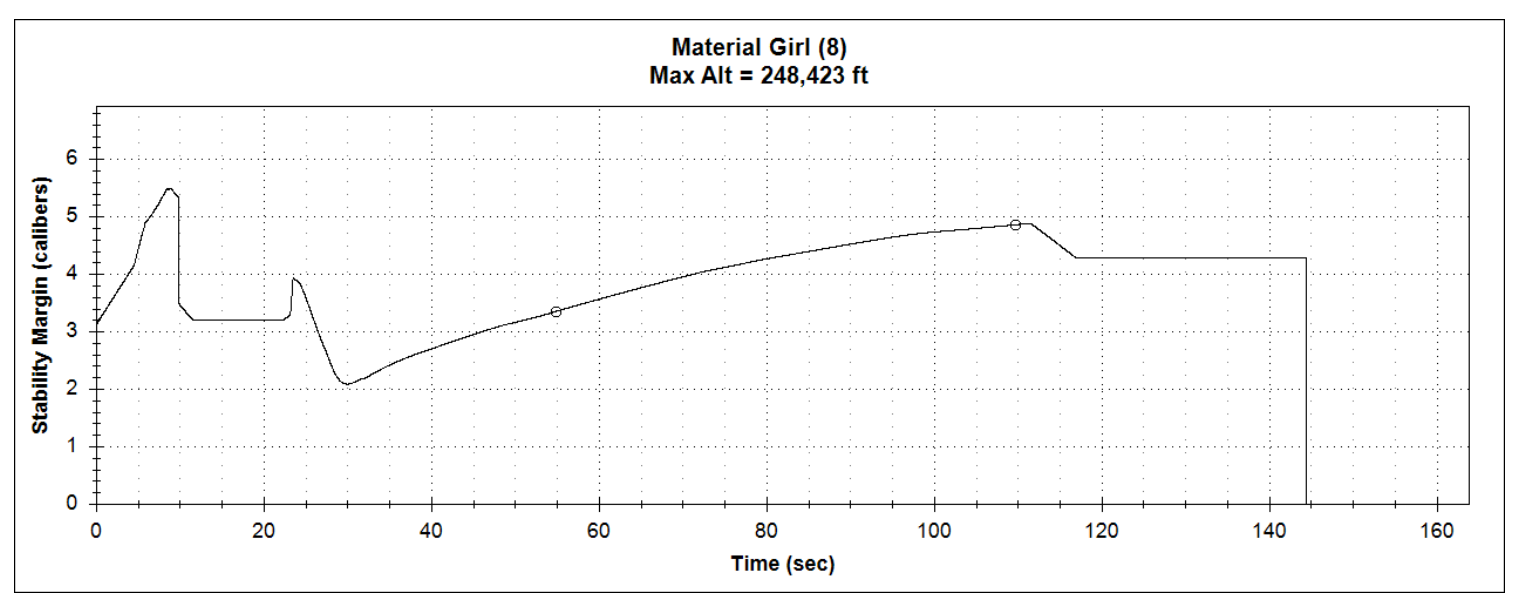}
    \caption{Stability Plot. Created using RASAero, with fin designs generated using an in-house genetic algorithm. The spike in stability margin around 10 seconds and the dip in stability margin at 30 seconds are unavoidable for a vehicle of this class. Stability margin increases rapidly initially due to the sheer mass of propellant expelled. The stability margin decreases as the vehicle reaches Max Q during the sustainer burn.}
    \label{fig:Stability}
\end{figure}

\subsubsection{Stability}
Dynamic stability was modeled using RASAero. RASAero is the baseline for computing
stability in the transonic and supersonic regime of high-power rocketry. It has proven effective
across multiple high-altitude solid rocket flights for modeling dynamic stability, and simulation
predictions have been validated against measured flight data. All stability analysis has been
based on this tool. Figure \ref{fig:Stability} shows the stability of \textit{Material Girl} during the duration of the simulated flight in RASAero.\\

The fins were designed using an in-house optimization script to meet strict stability constraints and fin flutter constraints. Four main stability constraints were applied: the booster starting stability must be above three calibers with a high out-of-rail exit velocity, all velocities above Mach one must have a stability margin above two calibers, the start of the sustainer burn must be at or above two calibers, and the stability at maximum dynamic pressure must be above 2.25 calibers. The stability plot shown in Figure \ref{fig:Stability} meets all four stability constraints while maintaining an advantageous drag and weight for obtaining the desired apogee. The stability modeling approach and the applied constraints described above have been demonstrated to produce nominal, low tilt, adequately stable, and safe flight profiles for three similarly sized vehicles. \textit{Material Girl} used the same process but with additional refinements and greater accuracy of mass, geometry, and thrust estimates.\\

\subsubsection{Dispersion Analysis}

A Monte Carlo-based dispersion analysis simulation was conducted with a sample size of 3,000 rockets. This simulation was run using the 6-DOF flight simulation library RocketPy. This dispersion analysis also included possible failure scenarios, including sustainer deployment failure and sustainer ignition failure. Each respective uncertainty parameter was given a reasonably large standard deviation to ensure that the dispersion data showcased a wide range of launch conditions. The basis for each of these values came from past flight experience, measurements taken from \textit{Material Girl}, and comparable data from similar vehicles. Improvements made to these estimates reflect the minimal expected design variability remaining in the overall vehicle and an understanding of past vehicles flown by GTXR. In addition to the uncertainty parameters, an ensemble of ten real-world, regional weather conditions was chosen at random for each rocket simulation; the weather data was taken from netCDF data sets. Real-world weather data coupled with the most accurate vehicle parameters possible provided the highest fidelity dispersion analysis our team could achieve. The full results of the dispersion analysis for this vehicle can be found in Appendix A. Figure \ref{fig:ApogeeFreq} below shows the apogee frequency of the dispersion analysis conducted. As expected, the mean apogee was approximately 220,000 feet. It is important to note that on the high end of this apogee curve, most rockets become infeasible, experiencing a large thrust scaling boost with a large mass reduction by chance of the randomness of the dispersion analysis program. Given the vehicle’s mass and thrust are well characterized, these vehicles tested in the analysis reflect an unreachable worst-case scenario by a significant margin and still landed in a safe vicinity of the launch area in both ballistic impact and high winds.\\

\vfill

\begin{figure}[h!]
    \centering
    \includegraphics[width=0.9\textwidth]{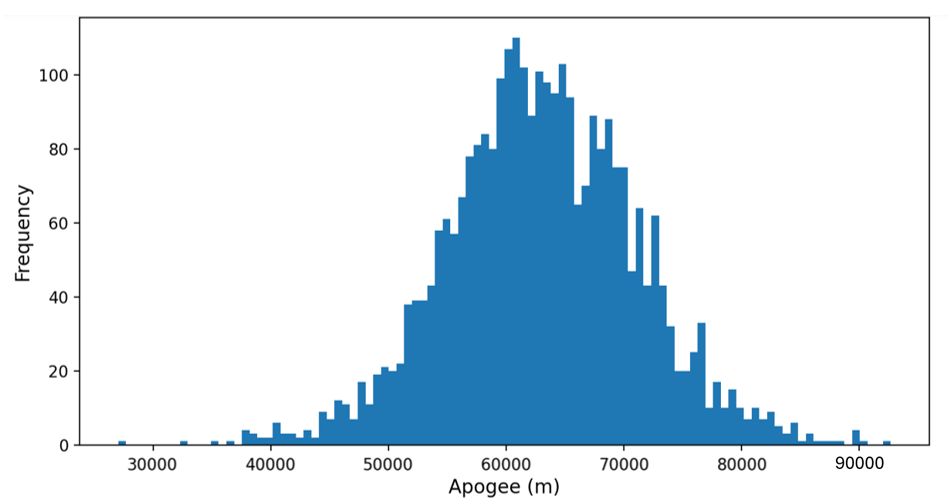}
    \caption{Apogee frequency plot for 3,000 rocket dispersion analysis.}
    \label{fig:ApogeeFreq}
\end{figure}

\vfill

\newpage

\subsection{Structures}

\textit{Material Girl}’s airframe was constructed out of filament-wound carbon fiber tubing with a high-temperature epoxy resin. The resin had a glass transition temperature of 400°F and a service temperature of 350°F to ensure the airframe remained operational at flight temperatures. These tubes were custom manufactured for our team by Rock West Composites to our specifications to ensure sufficient strength, stiffness, temperature resistance, and uniformity of the composite for our application. The tip of the nose cone was machined out of 316 stainless steel to resist deformation due to aerodynamic heating in flight. The structural considerations of the staging, recovery, and deployment systems are discussed at length in this section. Table \ref{table:MassLength} below shows the mass and length breakdown of each section of the vehicle.

\begin{table}[h!]
\centering
\caption{\textit{Material Girl}’s mass and length breakdown.}
\begin{tabular}{||c c c||} 
 \hline
 \textbf{Vehicle Section} & \textbf{Liftoff Mass (lbs)} & \textbf{Length (in)}\\[0.5ex] 
 \hline\hline
 Nose Cone & 7.59 & 36.5\\ 
 Sustainer &97.3 &74.0\\
 Booster & 92.2 &65.3\\
 \hline
\end{tabular}
\label{table:MassLength}
\end{table}

\subsubsection{Fins}

The fins of both the sustainer and booster utilized composite sandwich construction. The composite sandwich was made out of carbon fiber face sheets with a G10 fiberglass core. The fins were manufactured and cured individually before being bonded onto the airframe with Proline 4500 epoxy. This epoxy has a 275°F service temperature to provide the necessary rigidity for additional curing. Fillets of 0.75 inch radius were added to the seam between the root chord and the airframe also using Proline 4500 epoxy. These fillets provided substantial strength to the root chord-airframe bond as well as a rounded, aerodynamic surface to act as a form for the subsequent layers of carbon fiber. An eight-layer carbon fiber layup was then added to further adhere the fins to the airframe. This carbon fiber layup ran from the tip of each fin to the tip of the adjacent fin to ensure sufficient joint stiffness.\\

Fin flutter was analyzed using an in-house developed model. The stiffness of a composite sandwich fin was conservatively estimated through a series of classical lamination theory calculations and flutter was then predicted using the method described in NACA Technical Note 4197. The fin design optimizer used this model to select fin geometry as well as the number of carbon fiber layers. A factor of safety of at least two was required for the computed flutter Mach number compared against the flight Mach number at every simulated time step. The fin leading edges and nose cone were coated with Cotronics 4460 high-temperature epoxy to prevent delamination of the carbon fiber on the fins at flight temperatures. \\

\subsubsection{Shrouds}
The sustainer section of \textit{Material Girl} also had two externally mounted fiberglass shrouds that served as housing for the Featherweight GPS modules. This was required due to the carbon fiber airframe, which is opaque to radio frequency transmissions. These shrouds were bonded to the airframe using Cotronics 4460 high-temperature epoxy. This epoxy has been shown to remain structural at temperatures of 300 degrees Fahrenheit. The bond area between the shroud and the airframe was maximized to ensure sufficient strength. The aerodynamic loading in flight was estimated with shock-expansion theory, and the maximum load value was compared against the theoretical shear strength of the bond. A double-digit factor of safety was applied to account for any errors in load estimation or bond quality.\\

\subsubsection{Internal Hardware}

The internal structural components, including thrust rings and parachute mounts, were machined out of aluminum 6061-T6. Additional structural components were manufactured with G10 fiberglass. These materials were selected for their high specific strength and stiffness. Structural components were analyzed using finite element analysis. They were subsequently tested on an Instron load frame to ensure functionality at flight loads.\\

\subsubsection{Staging}
The booster and sustainer were rigidly connected during boost through a mechanical staging system utilizing a Marman band clamp. Symmetric flanges consisting of a lip and groove mated the booster to the sustainer. Thirty-two v-shaped blocks were bolted to a band of 0.020” thick spring steel wrapped around the flanges, securing the two stages together. Paracord ran through the ends of the band clamp, into the vehicle, and through two TinderRocketry Mako Line Cutters before being secured in a tensioning block. The resulting preload in the band clamped the two flanges together, providing the necessary force to resist bending and axial loads during flight. After booster burnout, the line cutters are actuated, cutting the tensioning cord. This releases the band from the flanges, allowing the two stages to drop apart.\\

The Mako Line Cutters use a small black powder charge to actuate a cutting piston. Two Mako Line Cutters were used for redundancy. These line cutters are hermetically sealed and have been tested successfully in a vacuum environment. The mechanical staging system has been tested to expected flight loads in bending and analyzed to flight loads in compression using finite element analysis. This process of analysis has shown that the staging mechanism will handle the loads expected in flight even under high angles of attack and velocities within standard safety margins. Additionally, this staging mechanism flew successfully on \textit{Mr. Blue Sky}, a similarly-sized vehicle, in the summer of 2022.\\

\subsubsection{Deployment}
The sustainer of \textit{Material Girl} had a parachute designed to deploy at apogee. The sustainer and nose cone parachutes were deployed using 7 grams of boron-potassium nitrate (BPN). BPN has been used successfully by GTXR on previous flights. It has been shown to ignite regardless of atmospheric pressure through tests conducted in a vacuum chamber. The sustainer and nose cone sections were joined together using three aluminum blind rivets. At the initiation of the deployment charge, the parachute bay of the rocket is pressurized with hot combustion gasses, at which time these rivets shear and allow for separation of the rocket sections. Parachutes have been shown to deploy successfully on multiple ground tests as well as during two separate drop tests conducted from a plane flying at low altitude. To ensure sufficient redundancy and reliability of deployment, \textit{Material Girl} had two charge wells, each with 7 grams of BPN, and each charge well has two redundant e-matches. Additionally, one e-match from each charge well is wired to each flight computer. This provides two layers of redundancy across all possible failure methods. Only one charge well ignition was required in order to deploy the parachutes. The booster utilized a passively deployed streamer released from the vehicle during staging.\\

\subsubsection{Recovery Systems}
\textit{Material Girl} used a drogue-only recovery scheme to eliminate the need for main parachutes and minimize drift distance. The vehicle was recovered in three separate sections: the booster section, the sustainer section, and the nose cone section. The nose cone section used a three-foot diameter, ultra-light, ripstop nylon parabolic parachute. The sustainer section used an ultra-high strength, five-foot, ripstop nylon parabolic parachute. The booster used a 50-foot-long ultra-high strength streamer. High-strength, 5⁄8” thickness tubular nylon shock cord connected each recovery device to its respective section of the rocket. Shock cords were protected from deployment gas with Kevlar sleeves that were installed over the shock cord.\\

Due to the projected deployment altitude and low air density of \textit{Material Girl}’s flight profile, the estimated snatch force at deployment was negligible. However, the parachutes impart a non-trivial force on the internal structure during descent. RASAero was used to find the descent acceleration of each section to extrapolate the tension in the shock cord. The internal structures of the vehicle were designed to these force values with appropriate factors of safety. The strength of these components was validated in an Instron load frame. In addition, the velocities were compiled across a range of possible deployment scenarios including deployment during staging and off-nominal deployment. All parachutes and shock cord were manufactured by Rocketman Parachutes. The nose cone used a Pro-X ultra-light drogue parachute and a 10-foot shock cord. The sustainer used a Ballistic Mach II parachute and two 30-foot shock cords, connected through a swivel. The booster used an Extreme Streamer with a 10-foot shock cord. Shock cords were joined to the vehicle and to recovery devices with 4500-pound-strength Kevlar soft links that have also been validated through Instron testing.\\

In addition to the parachutes and shock cord(s), each section also contained a radio-direction-finding (RDF) beacon, functioning as a backup recovery system in the event of GPS failure. The radio beacons, built by Communications Specialists and L. L. Electronics, broadcast tones on a known unique frequency between 220 MHz and 225 MHz, which can be picked up using a directional radio receiver (also built by Communications Specialists) to find the direction to the section. This system has been verified to remain functional after ground impact during drop tests.\\

\newpage
\subsection{Propulsion}
\subsubsection{Motors}

\textit{Material Girl} utilized two student-developed P-class solid rocket motors. The motors were functionally the same, with identical propellant composition and grain geometry. The sustainer’s nozzle was designed for optimal expansion at 30,000 ft, while the booster’s nozzle was expanded for sea level. The propellant was ammonium-perchlorate based and mixed to a team-developed formulation. This formulation and propellant geometry have flight heritage on past GTXR rockets as well as six nominal and successful static firings validating the P-class impulse rating of the motor design. Both motors employ the same BAllistic Test and Evaluation System (BATES) grain geometry, using five, ten-inch long grains designed by a student-made in-house grain optimization code. The grains were cast in phenolic casting tubes and epoxied into phenolic thermal liners. The liner and grains were enclosed inside an aluminum 6061-T6 casing, which served as the pressure vessel. The casing was sealed with a pinned forward closure and a pinned nozzle assembly. Thirteen radial pins retained closures for both motors; pins were rated for 1800 psi, two times the maximum expected operating pressure (MEOP) for adequate safety margin. The design’s factor of safety was verified using finite element analysis. To accept the pressure vessels for flight, hydrostatic testing was performed at a pressure of 1.5X MEOP (1,350 PSI) for a duration of five minutes. Each forward closure was machined out of a single piece of aluminum 6061-T6, including NPT holes for monitoring the motor’s pressure during static fire as well as passing through wiring.\\

The booster motor used a multiple-component aft-end assembly, with a nozzle carrier, a graphite nozzle throat, and a nozzle extension. The nozzle carrier was made of aluminum 6061-T6 and held the nozzle insert in place. The carrier also featured a convergent forward face to mitigate slag build-up during the motor burn. The nozzle throat interfaced with and was retained by the nozzle carrier. It was composed of fine grain graphite. The graphite throat featured transition geometry from the converging to the diverging section of the nozzle, capable of withstanding burn-duration exposure of high-temperature exhaust gasses. The nozzle throat sealed with the nozzle carrier via three piston seal O-rings. Along the diverging section of the assembly was a 316 stainless steel nozzle extension further expanding the exhaust flow to optimal pressure conditions for sea level. Stainless steel was selected to endure the high temperatures of the propellant exhaust. The sustainer motor used a similar multi-component nozzle assembly. This nozzle assembly used a similar base infrastructure as the booster but instead sealed the pressure vessel with a separate pin ring. This pin ring, made of aluminum 6061-T6, interfaced with the pins and pressure vessel. The nozzle carrier was retained by the face of the ring, and sealed with the casing using two piston seal O-rings. This seal was tested repeatedly and validated to withstand the resultant combustion gases during firing. The diverging section of the nozzle assembly was made of 316 stainless steel, bolted to the nozzle carrier using fifteen vibration-resistant nuts and bolts. This interface was sealed with a face-seal O-ring. The graphite nozzle insert was largely identical to the booster, with three piston-seal O-rings. Both motors were successfully static fired prior to flight, and all hardware was reusable with the exception of the graphite throats. A measured thrust curve and a table of key values are shown below. Thrust data for the sustainer was scaled appropriately in flight simulations to account for the lower atmospheric pressure during flight.\\

\begin{table}[h!]
\centering
\caption{Motor characteristics as determined by data from static fire. }
\begin{tabular}{||c c||} 
 \hline
 \textbf{Quantity} & \textbf{Value}\\[0.5ex] 
 \hline\hline
 Peak Thrust & 2207 lbf\\
 Impulse & 10349 lbf*sec\\
 Specific Impulse & 219.6 sec\\
 \hline
\end{tabular}
\label{table:1}
\end{table}

\begin{figure}[h!]
    \centering
    \includegraphics[width=0.9\textwidth]{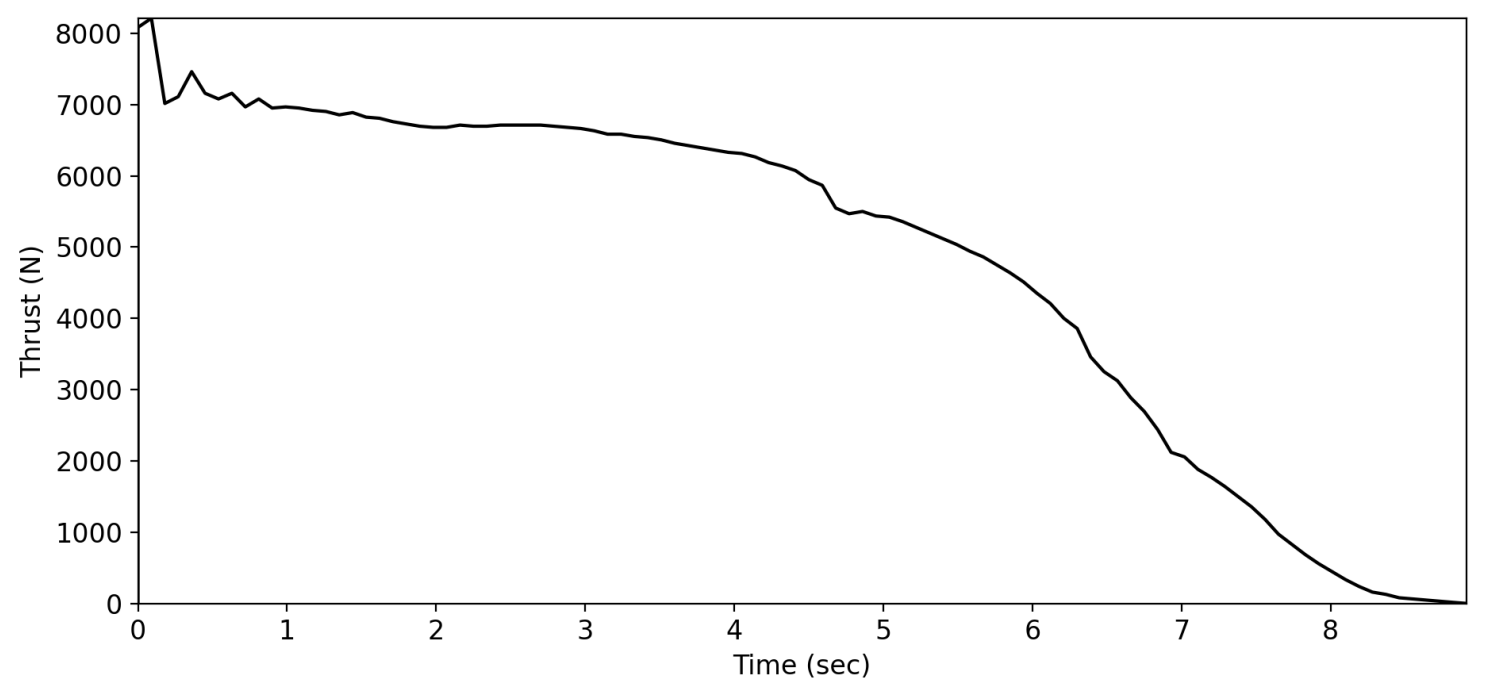}
    \caption{Thrust curve for booster and sustainer motor based on data acquired during the static fire.}
    \label{fig:image1}
\end{figure}

\subsubsection{Sustainer Ignition}
\textit{Material Girl} used a new head-end ignition system to increase the reliability of sustainer ignition. The igniter was a 20-gram, ammonium perchlorate propellant BATES grain cast into a 3D-printed casing. In the center of this grain, there was a powder mixture of boron potassium nitrate and propellant powder, produced as a by-product of grain post-processing. This powder was ignited using an e-match initiator. This two-stage igniter burned for approximately 10 seconds. Transient ignition time was about half a second when a current was supplied by a Raven V4 flight computer. The igniter threaded into the forward closure to secure it during flight. A pass-through in the forward closure connected the igniter wires to the flight computers. The passthrough was a Pave 1588-3, rated up to 2900 psi and 257°F. A thermal protection system constructed from a plate of G11 fiberglass, ceramic putty, and insulation protected the pass-through from the high temperatures of the motor.\\

As a redundancy measure, a legacy igniter design with flight heritage was installed alongside the main head-end ignition system. The legacy igniter consists of a commercial e-match surrounded by a small chunk of propellant and propellant powder. This ignition system was tested at the static fire of the sustainer motor. To ensure safety of all personnel, igniters were installed as late as possible in the assembly procedure to minimize the time and number of operations required with the sustainer igniter is in place. All assembly operations performed with the sustainer igniter installed were minimum personnel steps with extensive PPE. Additionally, three layers of protection existed to prevent premature firing of the sustainer igniters: switches that shunt the flight computer power and pyro batteries from the flight computer, a series of software interlocks on the flight computer, and an additional set of switches that shunt the sustainer igniters from their connection to the flight computer. None of these protections were moved to the armed state until \textit{Material Girl} was vertical on the pad.\\

\newpage

\subsection{Avionics}
\textit{Material Girl}'s avionics systems controlled the rocket's flight events (staging, sustainer ignition, and parachute deployment) as well as provided GPS location and onboard video recording. The subsystems performing these functions were distributed along three distinct avionics bays, one on each vehicle section that is independently recovered. The contents of the Nosecone (1), Sustainer (2), and Booster (3) bays will be discussed in the following sections.\\

\columnratio{0.2}
\begin{paracol}{2}
    
    \begin{figure}
    \centering
        \includegraphics[angle=-0.025, width=2.5cm]{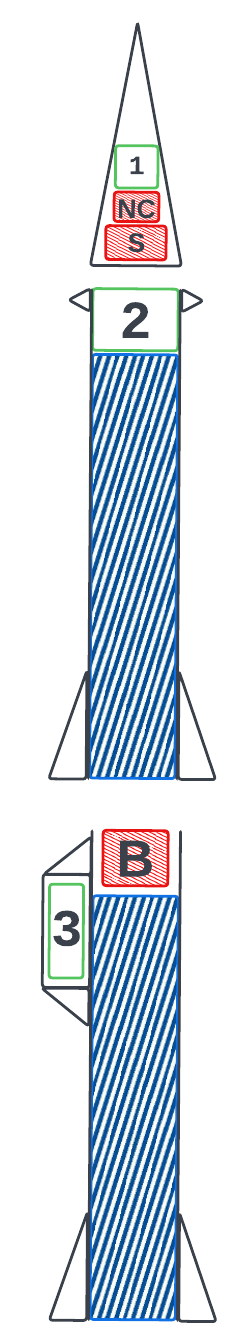} 
        \captionof{figure}{Distribution of avionics systems through \textit{Material Girl}. NS, S and B are the nosecone, sustainer, and booster parachute bays.}
        \label{fig:avBaysDiagram}
    \end{figure}
    \vfill
    
    \switchcolumn
    
    \subsubsection{Nosecone Bay}
        The Nosecone bay provided location capabilities to the vehicle's nosecone, which is detached from the sustainer after reaching apogee and deploying the sustainer parachute. It contains a Featherweight GPS module which transmits GPS location data through a LoRa connection. The bay is fully accessible from the outside of the vehicle via a conical nut that allows the bay to be completely removed from the rocket through the forward end of the nosecone. A render of this bay is provided in Figure \ref{fig:nosebayRender}.
        
    \begin{figure}[h!]
    \centering
    \includegraphics[angle=-90, width=4cm]{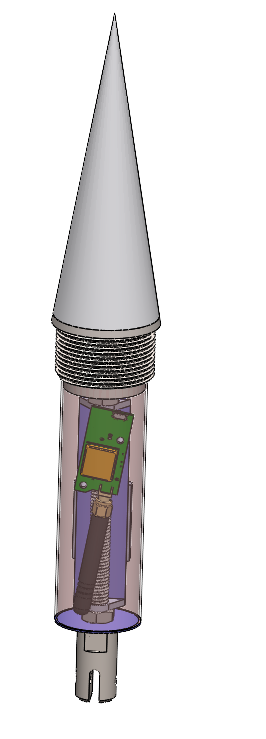} 
    \captionof{figure}{Nosecone Bay CAD rendering.}
    \label{fig:nosebayRender}
    \end{figure}
    
    \subsubsection{Sustainer Bay}
    The Sustainer Bay contained \textit{Material Girl}'s main flight computer, a Raven V4 developed by Featherweight Altimeters. The Raven altimeter was first released in 2009 and has over ten years of flight heritage. It uses integrated accelerometer data and barometer data to determine its state. The sustainer contained two Ravens for triggering flight events. Both Ravens were configured to perform sustainer ignition after a time delay from liftoff provided that an altitude of at least 4,000 ft had been reached. The apogee parachute was also triggered on a timer based on a nominal flight profile, with a redundant apogee charge using the barometer-based apogee detection within the Raven. Two Featherweight GPS modules were included in external shrouds and connected to this bay using SMA cables. These modules provided GPS location capabilities to this section of the vehicle. They were placed within shrouds to minimize the impact of the carbon fiber body tube on their LoRa antennas and maximize range at which their signal could be picked up by ground equipment. Finally, a custom flight computer being developed by the team was flown as a payload directly underneath the Raven V4. This flight computer did not command any flight events or transmit data, but was flown to develop the team's expertise in building and operating custom avionics systems. Finally, the bay contained two RunCam cameras, one pointing radially outwards from the vehicle and one pointing axially upwards into the parachute bay to observe parachute deployment. RunCam cameras were selected for their long battery life, long recording life, small form factor, and high resistance to the elevated ambient temperatures expected in the desert. The sustainer motor was ignited through a pass-through which allowed an electrical connection from the Raven flight computer to the motor igniter through the motor bulkhead. As the booster parachute was passively deployed, this bay configuration allowed for a single flight computer to control the entirety of the vehicle through all phases of flight, which greatly reduced flight event planning complexity.\\
    
\end{paracol}

\columnratio{0.6}
\begin{paracol}{2}

\subsubsection{Booster Bay}
The Booster Bay, nicknamed ``Muffin" for its peculiar shape, housed all avionics systems flying on the booster section of the vehicle. This included a Featherweight GPS system to provide location capabilities during recovery, a Raven V4 Flight Computer acting as a flight data logger and two ESP32-based camera boards developed by the team to collect video of the flight. The cameras were pointed along the axial direction of the vehicle and recorded through a thermo-formed transparent aerodynamic shroud, which protected the avionics systems from aerodynamic forces. This bay was entirely contained in a cup-shaped bay which was externally bolted to the vehicle, allowing teams easy access to the system while minimizing integration complexity.\\

\switchcolumn
\vfill
\begin{figure}
    \centering
    \includegraphics[width=0.15\textwidth]{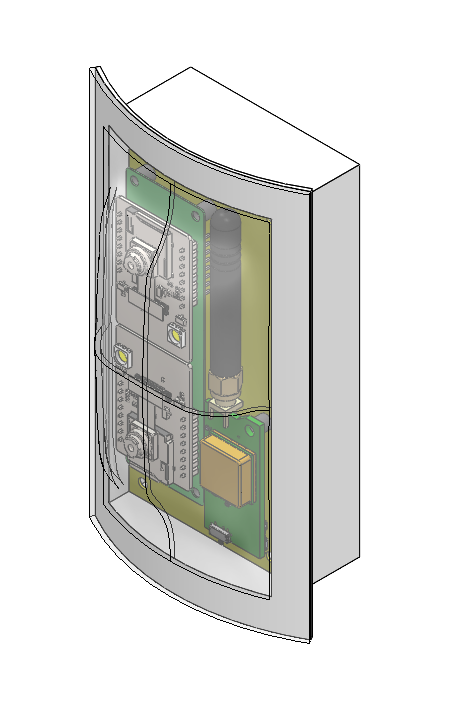}
    \label{fig:muffinShroud}
    \caption{Booster Bay CAD rendering.}
\end{figure}
\vfill
\end{paracol}

\section{Launch}
\subsection{Launch Timeline}
The launch preparation of \textit{Material Girl} was spread over two days with three separate assembly crews, named white, gold, and navy. The white team first arrived at FAR at 8 A.M. on Friday, July 7th and set to unpacking supplies and beginning full assembly without energetics in preparation for an All-Up test. The gold team replaced them at 2 P.M. to finish assembly and conduct the All-Up test, powering on all avionics and ensuring all systems behaved nominally. After successfully conducting the All-Up, the gold team began minimally disassembling the rocket to remove and recharge batteries and begin packing energetics. The navy team replaced them at 8 P.M. to assemble both rocket motors and the staging systems. The white team returned Saturday morning at 2 A.M. to complete full assembly and motor integration. The rest of the team then returned at 7 A.M. to install the sustainer motor igniter and join the two stages. Overall, integration ran incredibly smoothly, with all assembly crews finishing their shifts early. The only serious issue occurred during sustainer igniter integration, where the igniter assembly was discovered to be too large to fit through the nozzle.\\

 This was resolved by separating the assembly into its threaded propellant puck and auxiliary igniter. The two stages of \textit{Material Girl} were then joined and brought to the launch rail. When sliding the rocket on for the first time, the staging system snagged and lost tension, causing the two halves of the rocket to sag apart. The rocket was slid back off and the stages were separated. After re-tensioning the staging system and carefully sliding the rocket back on the rail, no further rigidity issues between the two stages were noticed. The pad was then cleared and the rail was raised vertical. A minimum personnel crew began powering on avionics and radios. GPS lock was successfully achieved with the avionics located in the nosecone and sustainer, but not the booster. Booster GPS lock was not expected to improve and was not a requirement for flight, so procedures continued. When checking e-match channels, it was found that the staging channel had no continuity, rendering the planned active stage separation impossible. After consulting go/no-go conditions, it was determined that staging continuity was not a requirement for flight under the condition of positive continuity on the sustainer motor igniter. Taking the rocket apart to attempt to fix this issue was deemed too time-consuming and dangerous, as it would involve a full disassembly of the sustainer motor. With this in mind, the pad crew removed the last switch shunting the sustainer igniters from the flight computer, and returned to the bunkers. Sustainer igniter continuity was achieved and the launch countdown continued per launch procedures with the expectation of a hot-staging during flight.\\

 \subsection{Relevant Procedures}

Below are the main points of the launch procedures and safeguards that are relevant to the arming and ignition of \textit{Material Girl}. Timelines and constraints below are reflective of past launches at FAR and past motor firings conducted by GTXR.\\

\subsubsection{Launch Countdown Procedure}
After being loaded onto the rail and the pad area has been cleared, the following arming and countdown procedure will be followed:
\begin{enumerate}
  \item Clear personnel from the pad.
  \item Power up and arm Ravens.
  \item Verify good radio link and GPS lock.
  \item Power up and arm payload flight computer.
  \item Verify e-match continuities.
  \item Pull sustainer shunt switches to arm sustainer igniters.
  \item Clear range.
  \item Ensure sky and range are clear to launch.
  \item Count down and initiate launch.
\end{enumerate} 

\subsubsection{Hang Fire Procedure}
In the event of the hang fire, the following procedure will be followed:
\begin{enumerate}
  \item Wait 30 seconds and retry booster ignition.
  \item If the rocket does not launch, disarm avionics.
  \item Wait five minutes before approaching the rocket.
  \item Send two people to approach the vehicle to safe, remove, then replace failed booster igniters. This practice will be conducted under careful scrutiny and in cooperation with FAR personnel.
  \item Return to bunkers, proceed with arming and countdown procedure.
\end{enumerate} 

\subsubsection{Launch Commit Criteria}
The following conditions must be met in order to launch:
\begin{itemize}
    \item Sky must be clear of aircraft.
    \item Visibility is sufficiently good to see approaching aircraft.
    \item The flightpath is free of clouds.
    \item No vehicles are approaching on the road.
    \item The area surrounding the launch rail is free of flammable materials.
    \item \textit{Material Girl} has been assembled correctly.
    \item GTXR technical go/no-go criteria have been satisfied.
    \item Friends of Amateur Rocketry have given go for launch authorization.
\end{itemize}

\subsection{Flight Timeline}
Launch was initiated by the FAR ground ignition system, connected to the igniter inside the booster motor. The booster motor ignited correctly, after which an anomaly in the sustainer parachute deployment system was experienced. This caused the vehicle nosecone to completely separate from the rest of the vehicle shortly after takeoff. The booster and sustainer continued to fly for approximately 5 seconds, after which a second anomaly was experienced with sustainer ignition. The sustainer and booster continued to fly separately until both motors burned out, after which no successful parachute deployment was observed. Both vehicle sections are presumed to have fallen back under a ballistic trajectory, after which they were located and recovered by the team. 

\subsection{Data Collected}
The anomaly experienced by the vehicle compromised the recovery of flight data from avionics subsystems on the vehicle that did not have data transmit capabilities. This includes the primary Raven flight computer and the custom flight computer payload. However, GPS flight data from the sustainer was recorded by the onboard Featherweight GPS Tracker.  After both stages returned to Earth, recovery crews were sent to locate the booster, booster avionics bay, and sustainer.  The trajectory data and recovery maps are found below. 

\begin{figure}[h!]
    \centering
    \includegraphics[width=0.9\textwidth]{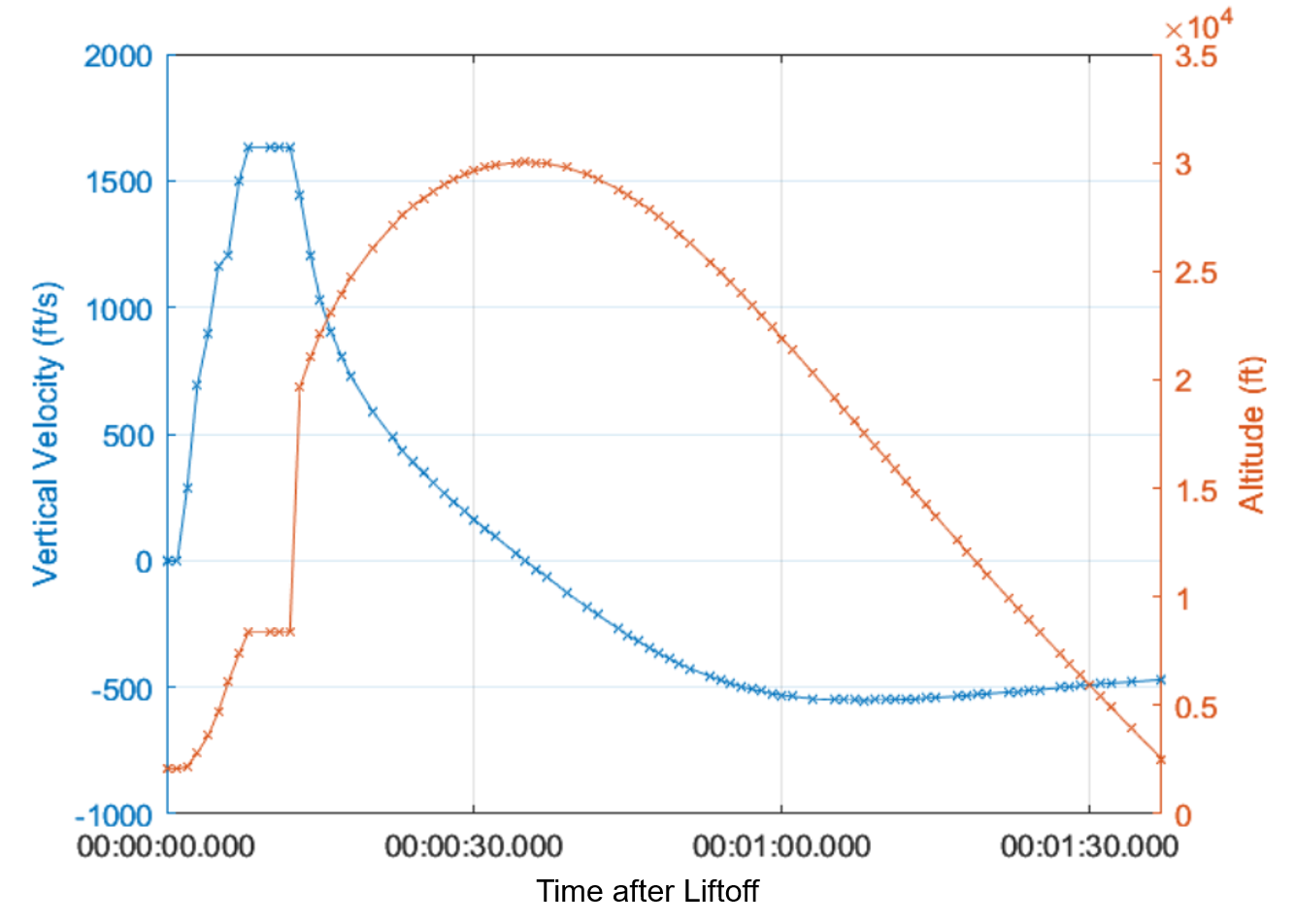}
    \caption{Sustainer Flight Trajectory.}
    \label{fig:susFlight}
\end{figure}

\begin{figure}[h!]
    \centering
    \includegraphics[width=0.5\textwidth]{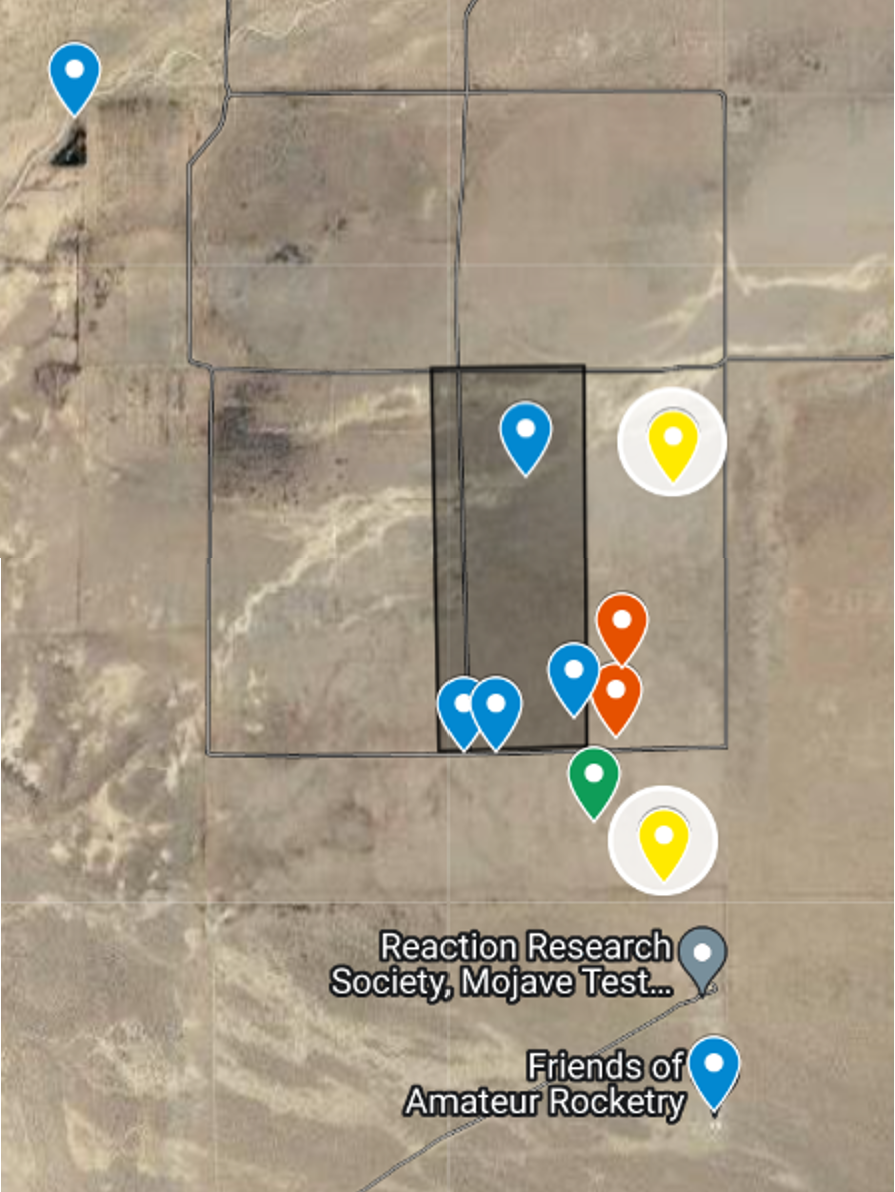}
    \caption{Sustainer Recovery Location (top circle) and Booster Recovery Location (bottom circle).}
    \label{fig:image1}
\end{figure}

\begin{figure}[h!]
    \centering
    \includegraphics[width=0.5\textwidth]{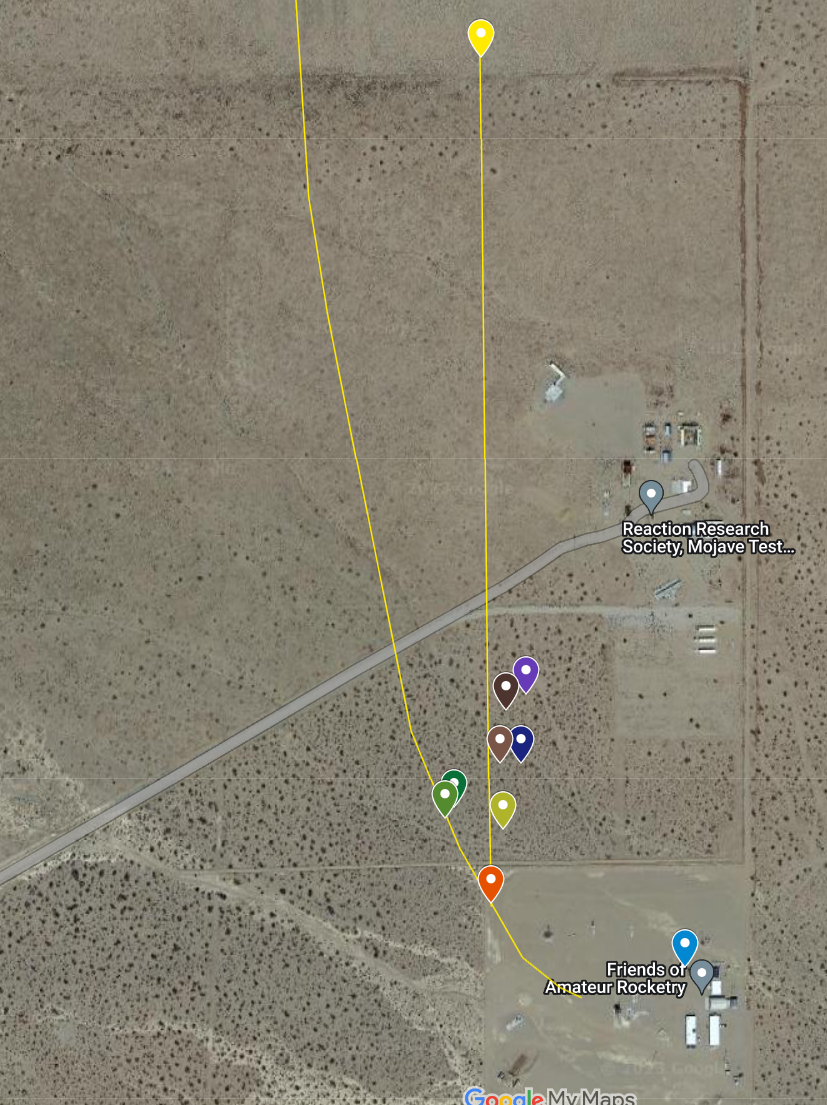}
    \caption{Booster Avionics Bay Recovery Location (yellow marker).}
    \label{fig:image1}
\end{figure}


\clearpage
\newpage
\section{Accident Investigation}

Upon launch, the vehicle faced a flight computer anomaly, which caused the nose cone to deploy from the vehicle while on the rail, followed by the sustainer hot-staging from the booster shortly after. Both stages reached an apogee of approximately 30,000 ft AGL before ballistically falling to the ground. An accident investigation was opened by the team to understand the conditions leading up to the anomaly and subsequent suboptimal flight to mitigate the risk of other anomalies impacting future GTXR vehicles.\\

\subsection{Flight Event Analysis}
The team used photos taken by GTXR alumn Casey Wilson during the launch in order to construct a timeline of events. Images showing relevant flight events are shown below, while an additional, more extensive, image gallery with annotated commentary can be found in Appendix B.\\

\begin{figure}[h!]
    \centering
    \includegraphics[width=0.9\textwidth]{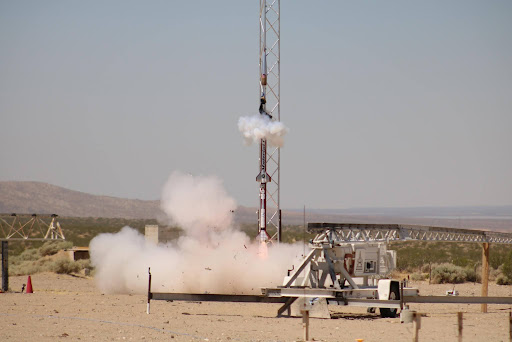}
    \caption{Nose cone separation (T+0:01)}
    \label{fig:image1}
\end{figure}

At T+0:01, the nosecone parachute charge well was triggered prematurely, resulting in the nosecone separating from the rest of the vehicle while still on the rail. This deployment caused the forward end of the sustainer to be exposed for the majority of the flight in an anomalous configuration.
\newpage
\begin{figure}[h!]
    \centering
    \includegraphics[width=0.9\textwidth]{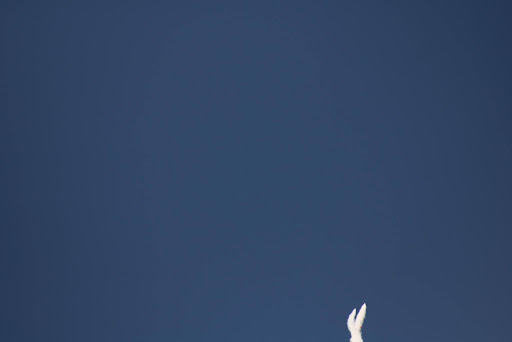}
    \caption{Sustainer Separates (T+0:05)}
    \label{fig:image1}
\end{figure}

At T+0:05, the second stage igniter was triggered prematurely, resulting in an early hot-staging event. This caused the sustainer to fire significantly before intended and while the booster was still accelerating. The violent hot-staging caused the booster recovery system to be inoperable and ejected the first-stage avionics bay. This ejection was captured by the instrumentation within this bay and the data was successfully recovered.\\

Images and video from the launch as well as data received from the GPS trackers located along the vehicle and the flight configuration used to program the Blue Raven Flight Computer were used to reconstruct the complete sequence of events leading up to and after the two anomalies experienced during flight.\\

\subsection{Anomaly Explanation}
Investigation into flight events indicated a software issue in the flight computer which caused all channels to fire prematurely.  The flight time of the vehicle was initialized when the flight computers were powered on instead of at launch detection. Because significant time had elapsed between arming and launch, the timer-based nosecone deployment channel was fired as soon as launch was initiated and boost was detected. Upon reaching an altitude of 4,000 ft AGL, the firing conditions for the sustainer igniter were met. This combination of events resulted in the early hotstaging of the second stage.\\

\subsection{Future Mitigation Plan}
In order to prevent similar events from occurring in future flights, the team has developed a set of comprehensive testing procedures to mimic flight environments and timelines as closely as possible on the ground. These tests will subject the flight computer to a realistic arming timeline and boost loads. The testing campaign will verify that the flight computer will perform nominally in a launch and flight environment. The team has also developed additional robust testing plans to qualify and proof various flight components to ensure that they are ready for future launches. These tests will supplement all of the tests conducted in the 2022-2023 academic year in the development of future vehicles.\\

\newpage
\section{Future Steps}
\subsection{\textit{Fire on High} and \textit{Strange Magic}}
After launch, GTXR performed a launch accident investigation and debrief in order to understand the events that unfolded. The team then conducted a series of architecture meetings to lay out the goals of the 2023-2024 academic year, deciding to design and fly two single-stage rockets. \textit{Strange Magic}  will test a new custom flight computer and internal systems designed to support it. It will serve as a testbed to develop feasible and simple integration procedures for this new system to aid in the reuse of this flight computer in the spaceshot attempt. \textit{Fire on High} will test novel propulsion and external systems. This rocket will feature GTXR's first-ever eight-inch diameter motor, which will play an instrumental role in providing the necessary power to take the team to space. This rocket will serve as a test bed for fin and shroud solutions to tackle the structural and thermal challenges of high speeds. These rockets are slated to fly in the summer of 2024. The team reasoned that in order to progress various systems toward a spaceshot vehicle while minimizing risks and unknowns, it would be prudent to independently test various systems for spaceshot in separate rockets.\\

\subsection{Conclusion} 
The flight of \textit{Material Girl} on July 8th, 2023 was the culmination of GTXR's yearlong effort in building a two-stage sounding rocket.  Launch of the vehicle resulted in premature deployment and hot-staging due to a timing error in the flight computer, causing the sustainer to reach an apogee of 30,000 ft AGL.  Development and integrated testing of a custom flight computer is currently in progress, which will give the team greater confidence in the avionics systems of future vehicles.  In order to develop various technologies toward a spaceshot vehicle, GTXR will fly two single-stage rockets in Summer 2024.  \\ 

\section{Acknowledgements}
GTXR thanks the Georgia Tech Student Government Association, the Daniel Guggenheim School of Aerospace Engineering, the Georgia Space Grant Consortium, the Georgia Tech Honors Program, the Georgia Tech Foundation, and the Georgia Tech Student Foundation for their financial support of the launch. GTXR also thanks its team partner, Hexagon Manufacturing Intelligence.\\

Hexagon products were utilized in the design and analysis of many of \textit{Material Girl}'s structures and propulsion components. The ability to model composite parts accurately as well as perform finite element analysis on the vehicle's components was crucial in making this launch possible. These approaches will be further utilized in the design of GTXR's next high-altitude rocket, \textit{Fire on High}.\\

Furthermore, this year's accomplishments at the launch would not have been possible without the support of the team's corporate sponsors: Blue Origin, Rock West Composites, SpaceWorks, Delta Tech Ops, and Kenesto. The team would also like to thank City Auto Body Shop, Graphite Store, and Mr. Jonathan Shi for their support of this year's vehicle.\
\newpage
\appendix
\renewcommand{\thesection}{\hspace{-1em}}
\renewcommand{\thesubsection}{\Alph{subsection}}
\section{Supplementary Materials}
\subsection{Dispersion Analysis}

 Figure \ref{fig:BoosterRec} shows a nominal booster stage dispersion for the planned vehicle flight profile. As seen in Figure \ref{fig:SustainerRec} and Figure \ref{fig:NoseRec}, even with a generous standard deviation for launch conditions, the first and second standard deviation ellipses for the sustainer and nose impact sites lay comfortably within an 11-mile distance of the launch site (indicated by the red point), most impacts being much closer. The third standard deviation ellipses for both the sustainer and nose cone do not cross the 20-mile mark. The furthest impact site for the nose cone was approximately 17 miles from the launch site; however, the probability of such a launch occurring was extremely low and the result was still well within a safe landing area. The furthest impact site for a ballistic second stage with separation failure was approximately 75 miles from the launch site. As mentioned earlier, Figure \ref{fig:SustainerRecFailedSep} shows the worst-case scenario dispersion: failed deployment at the apogee of a nominal flight. While the vehicle will impact safely but still very far from the launch site, it is also worth noting the unlikelihood of this scenario.\\

\columnratio{0.5}
\begin{paracol}{2}
\begin{figure}[]
    \centering
    \includegraphics[width=0.45\textwidth]{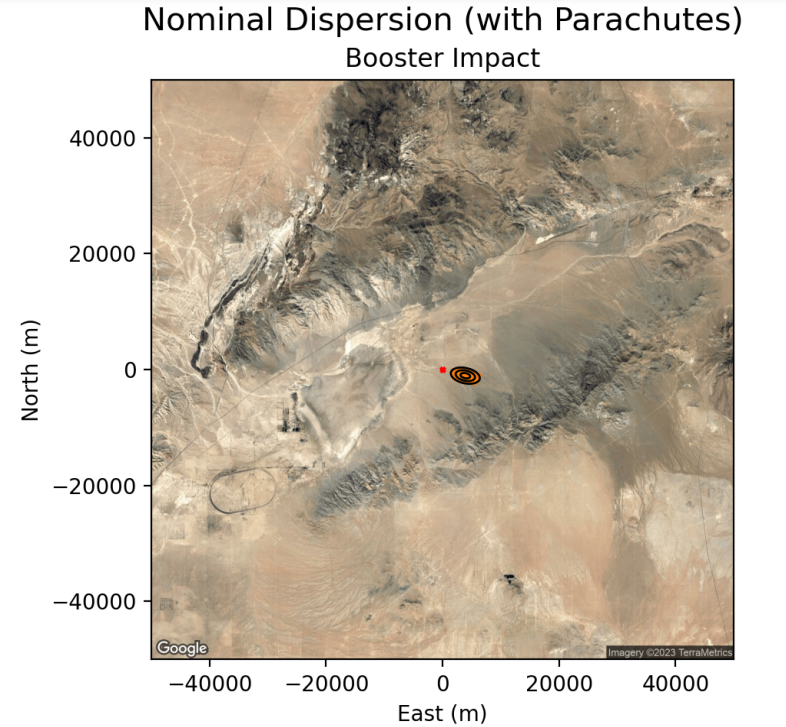}
    \caption{Dispersion of booster recovery flight event with standard deviation ellipses and launch site marked by red dot. As shown by the plot, the booster will land very close to the launch site.}
    \label{fig:BoosterRec}
\end{figure}
\switchcolumn
\begin{figure}[]
    \centering
    \includegraphics[width=0.45\textwidth]{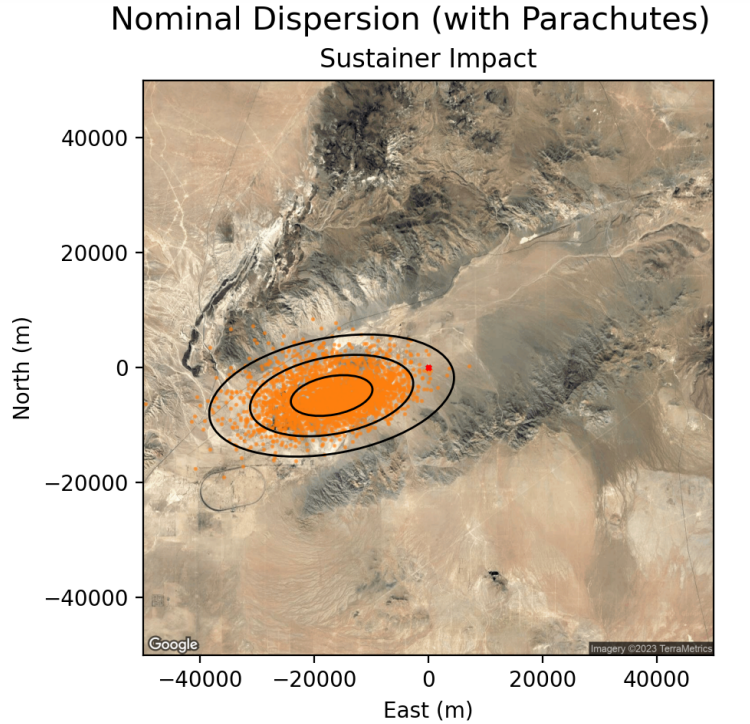}
    \caption{Dispersion of sustainer recovery flight event with standard deviation ellipses and launch site marked by red dot. As seen by the plot, the sustainer will nominally land within roughly 20 miles of the launch site over unpopulated areas.}
    \label{fig:SustainerRec}
\end{figure}
\end{paracol}

\columnratio{0.5}
\begin{paracol}{2}
\begin{figure}[h!]
    \centering
    \includegraphics[width=0.4\textwidth]{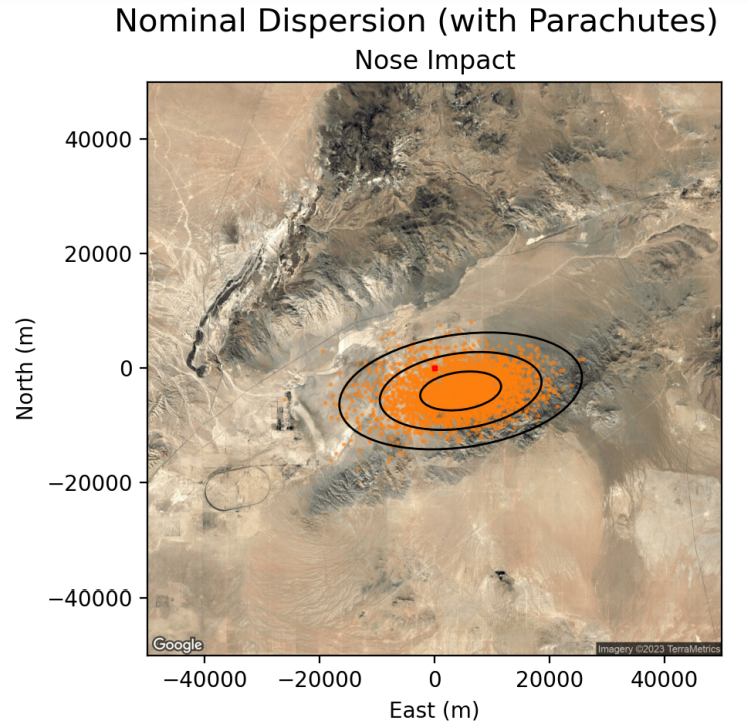}
    \caption{Dispersion of nose recovery flight event with standard deviation ellipses and launch site marked by red dot. As seen by the plot, the nose cone will nominally land within roughly 10 miles of the launch site over unpopulated areas.}
    \label{fig:NoseRec}
\end{figure}
\switchcolumn
\begin{figure}[h!]
    \centering
    \includegraphics[width=0.45\textwidth]{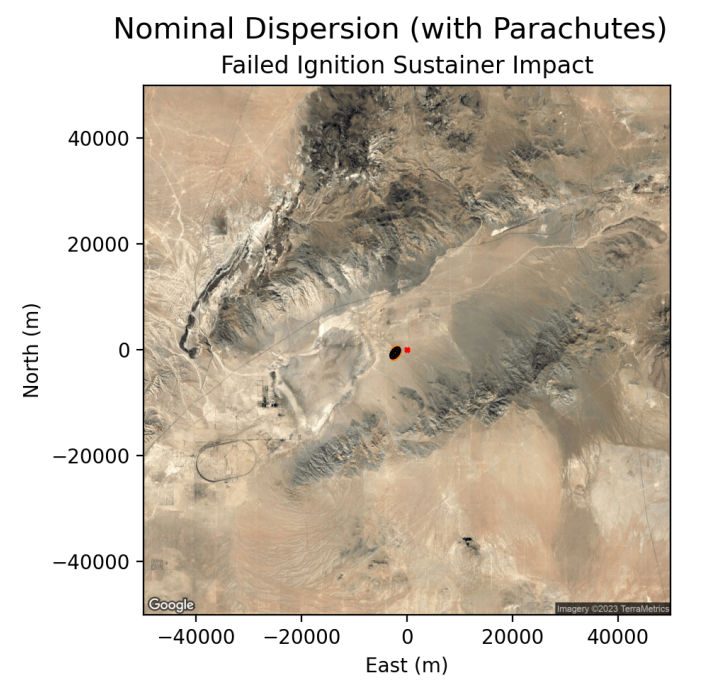}
    \caption{Dispersion of sustainer recovery flight event for a failed sustainer ignition with standard deviation ellipses and launch site marked by red dot. In the event of a failed sustainer ignition, the sustainer will land very close to the launch site.}
    \label{fig:SustainerRecFailedIgniter}
\end{figure}
\end{paracol}
\newpage

\columnratio{0.5}
\begin{paracol}{2}
\begin{figure}[h!]
    \centering
    \includegraphics[width=0.4\textwidth]{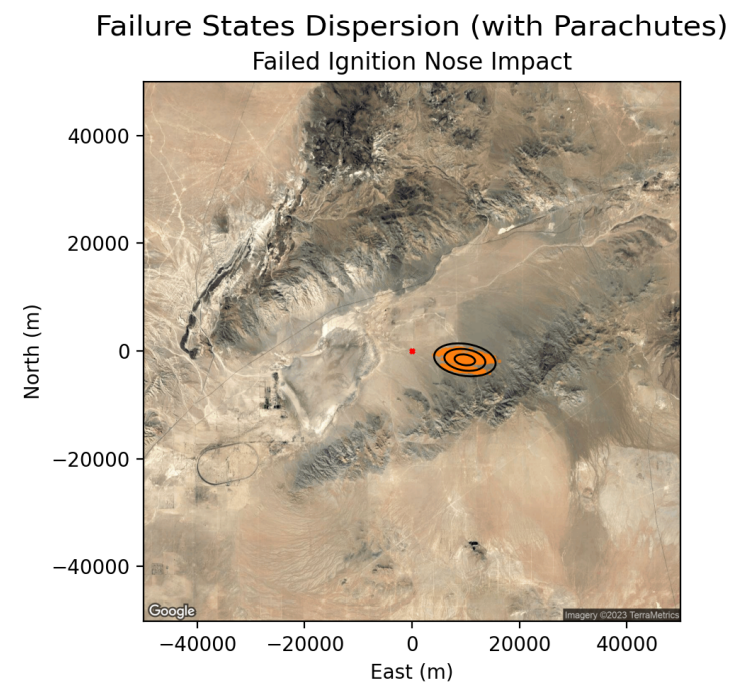}
    \caption{Dispersion of nose recovery flight event for a failed sustainer ignition with standard deviation ellipses and launch site marked by red dot. In the event of a failed sustainer ignition, the nose cone will land very close to the launch site.}
    \label{fig:NoseRecFailedIgniter}    
\end{figure}
\switchcolumn
\begin{figure}[]
    \centering
    \includegraphics[width=0.45\textwidth]{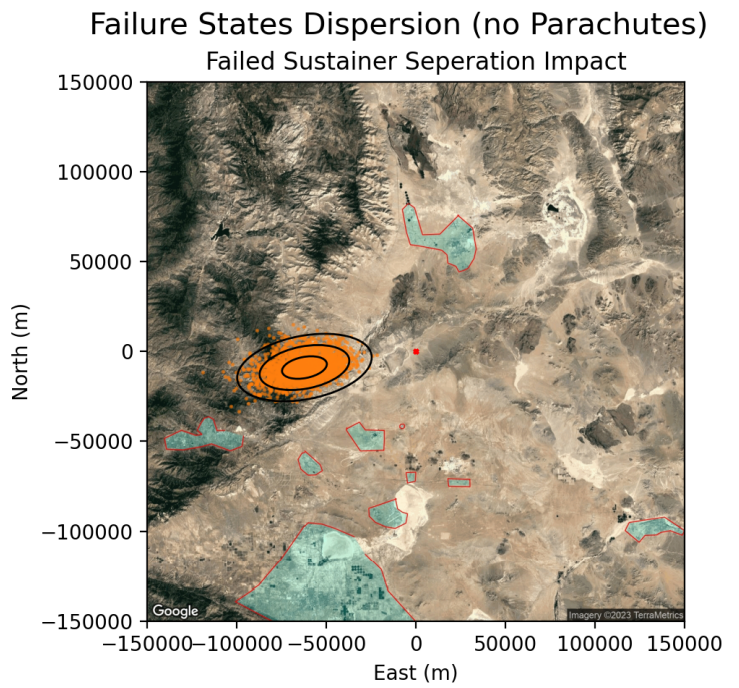}
    \caption{Dispersion of sustainer recovery flight event for a failed nose/sustainer separation with standard deviation ellipses and launch site marked by red dot. Populated areas are highlighted in blue. This is the most extreme case of dispersion. As shown in the plot, while the distance from the launch pad is considerable, the ballistic sustainer will still land well clear of populated areas even at three standard deviations from nominal conditions.}
    \label{fig:SustainerRecFailedSep}
\end{figure}
\end{paracol}
\clearpage
\subsection{Image Gallery of Flight Events}

This appendix contains annotated images taken during the first few minutes of \textit{Material Girl}'s launch. The team would like to thank Casey Wilson for putting together this analysis, which greatly helped in determining the series of events leading up to and right after the anomaly experienced during flight.\\

The camera parameters used to take these images are as follows:

\begin{itemize}
    \item Camera model: Canon T6s\\
    \item Time in between photos in continuous operation: ~0.2sec\\
    \item Shutter speed: 1/4000\\
    \item Sensitivity to light: ISO 200\\
    \item Focal length: F4.0\\
\end{itemize}

Timestamps on the images are specified to 1-second accuracy.

\begin{figure}[h!]
    \centering
    \includegraphics[width=0.9\textwidth]{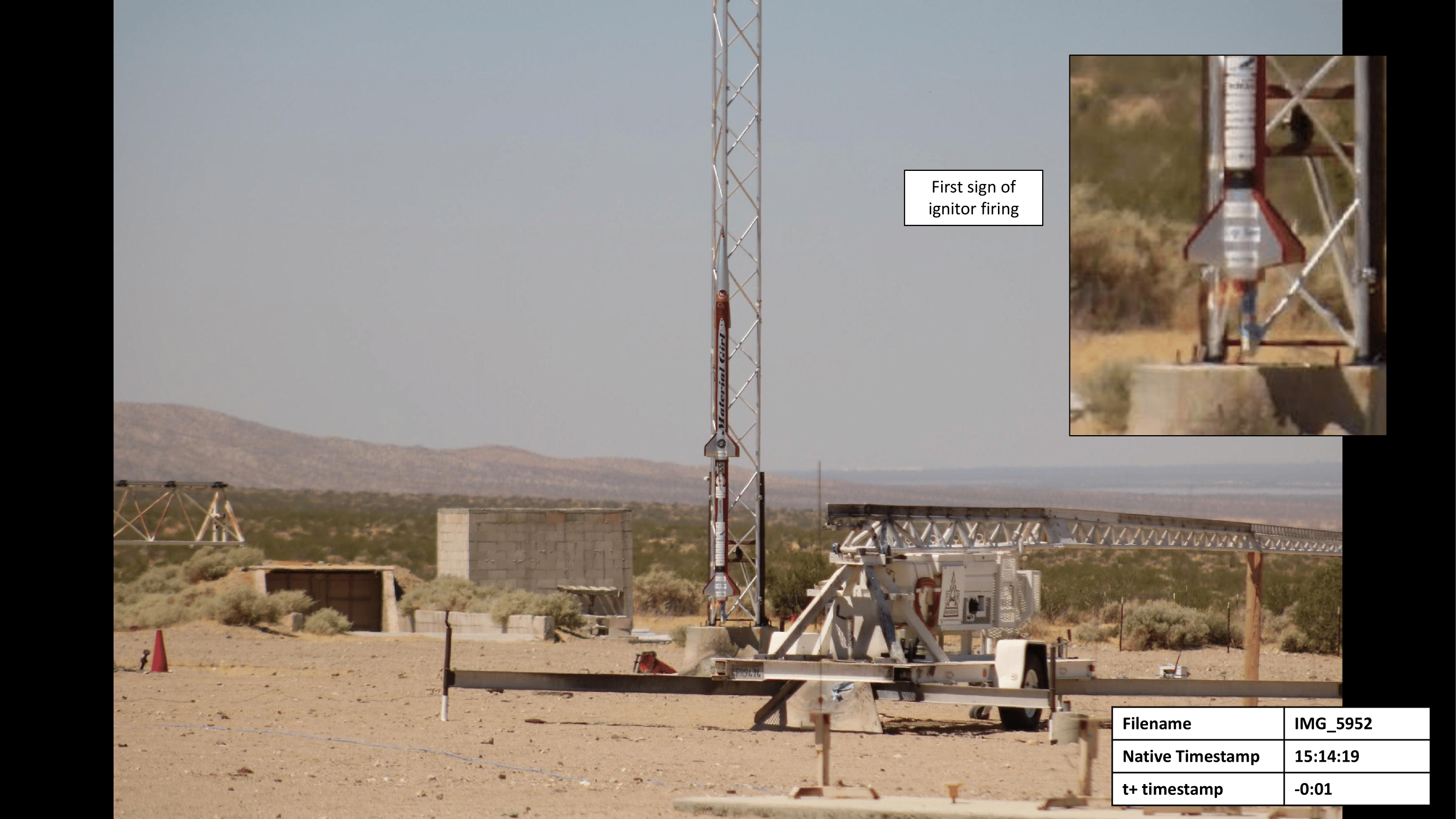}
\end{figure}

\begin{figure}[h!]
    \centering
    \includegraphics[width=0.9\textwidth]{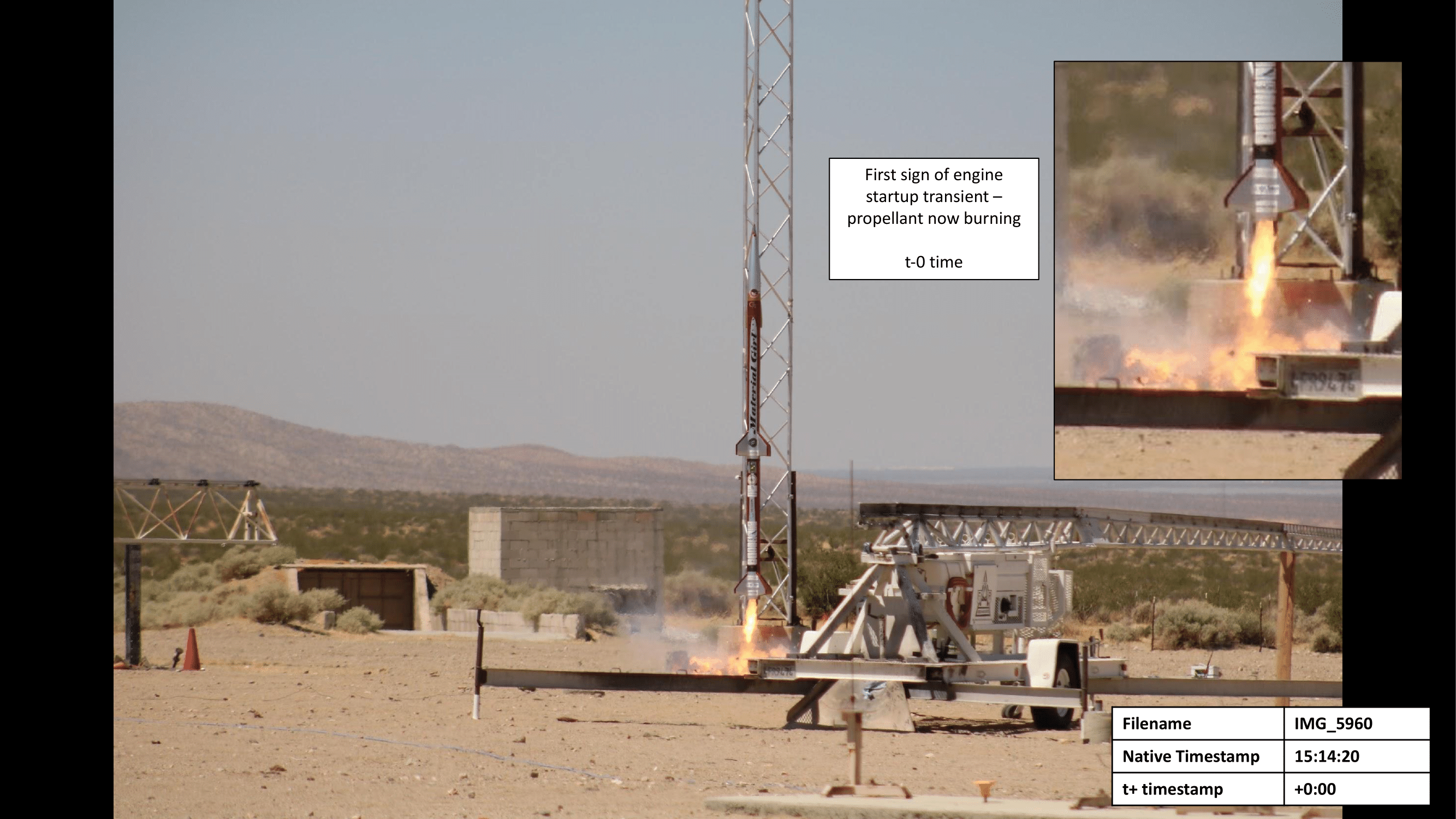}
\end{figure}

\begin{figure}[h!]
    \centering
    \includegraphics[width=0.9\textwidth]{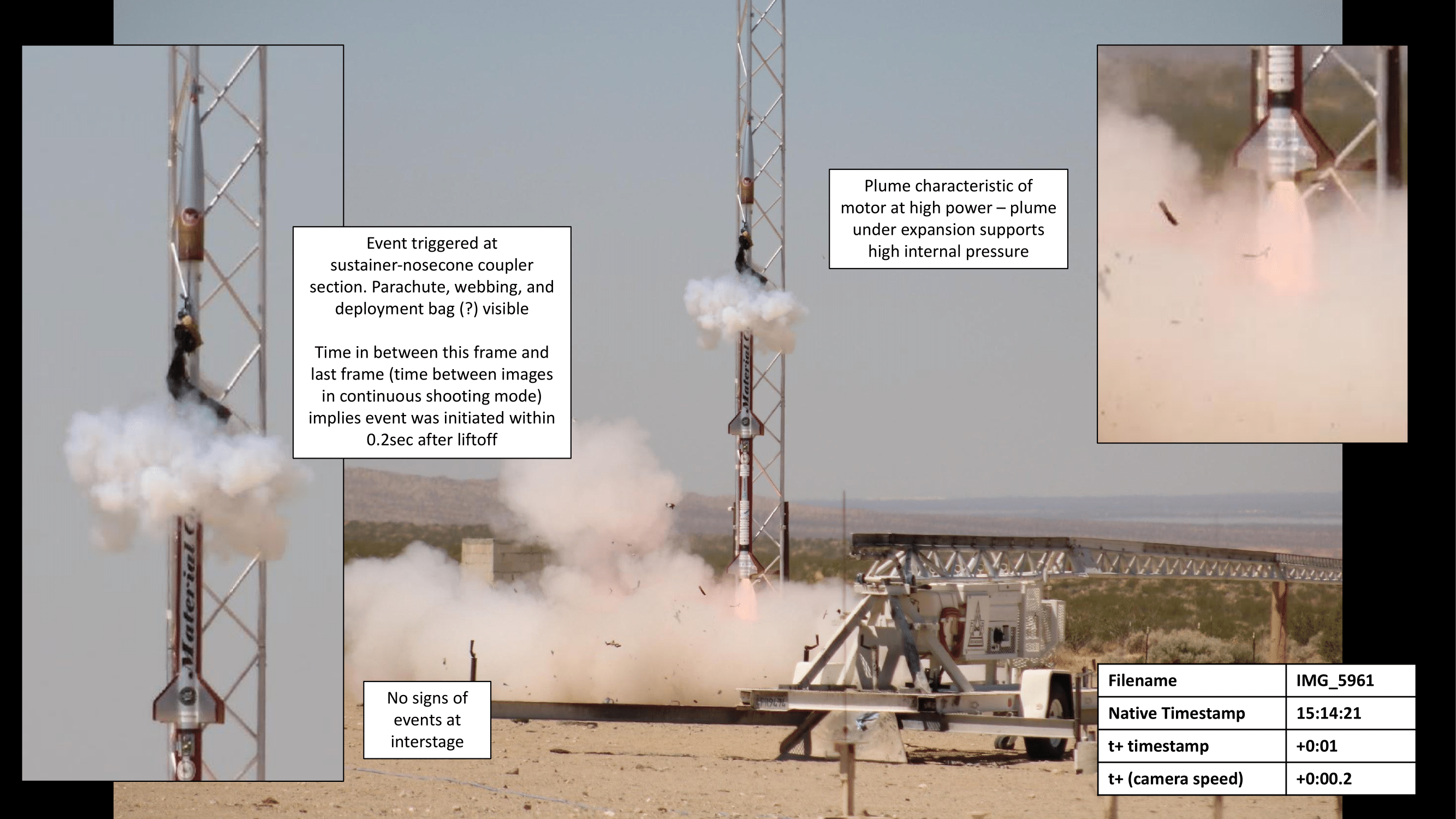}
\end{figure}

\begin{figure}[h!]
    \centering
    \includegraphics[width=0.9\textwidth]{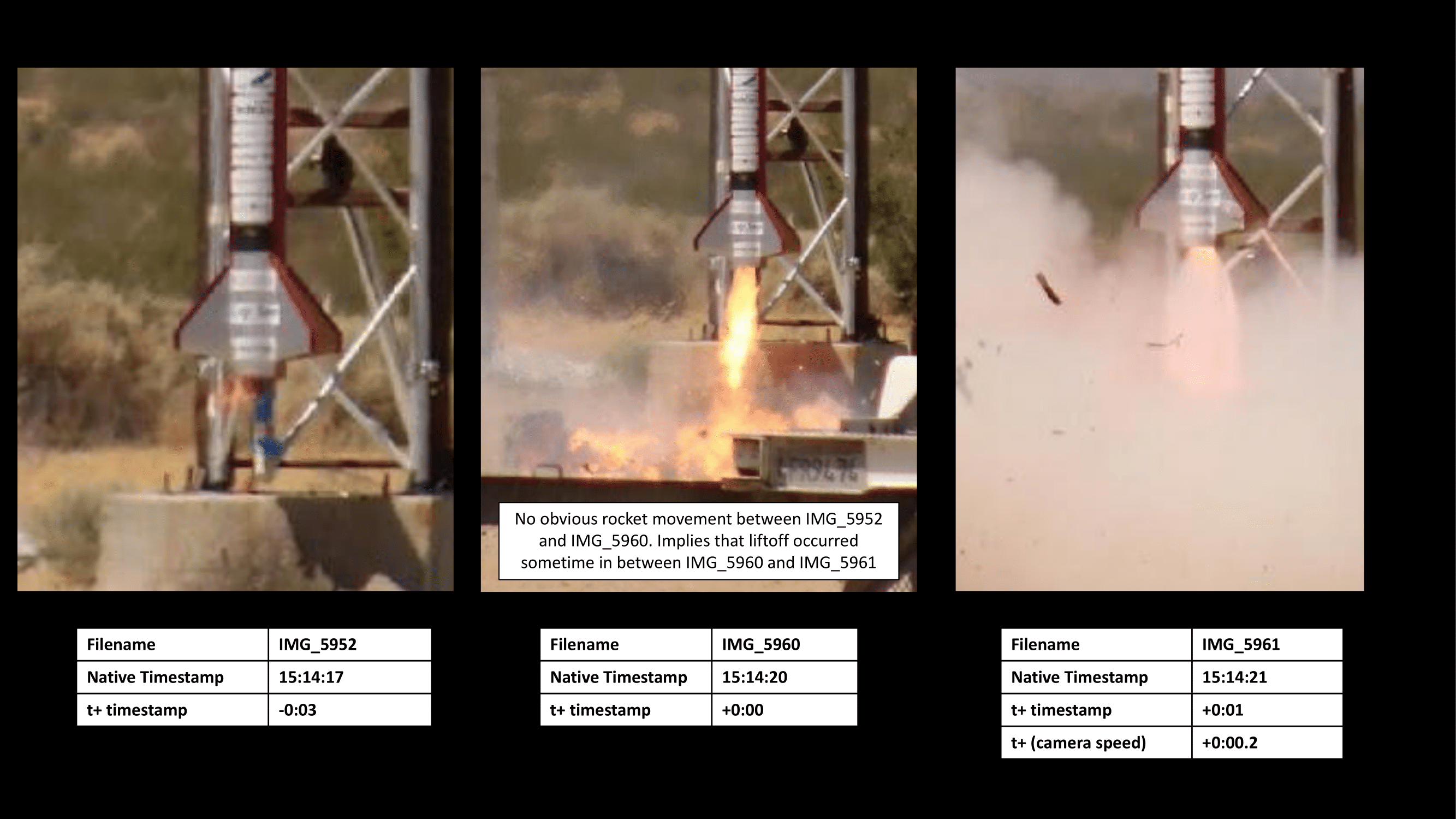}
\end{figure}

\begin{figure}[h!]
    \centering
    \includegraphics[width=0.9\textwidth]{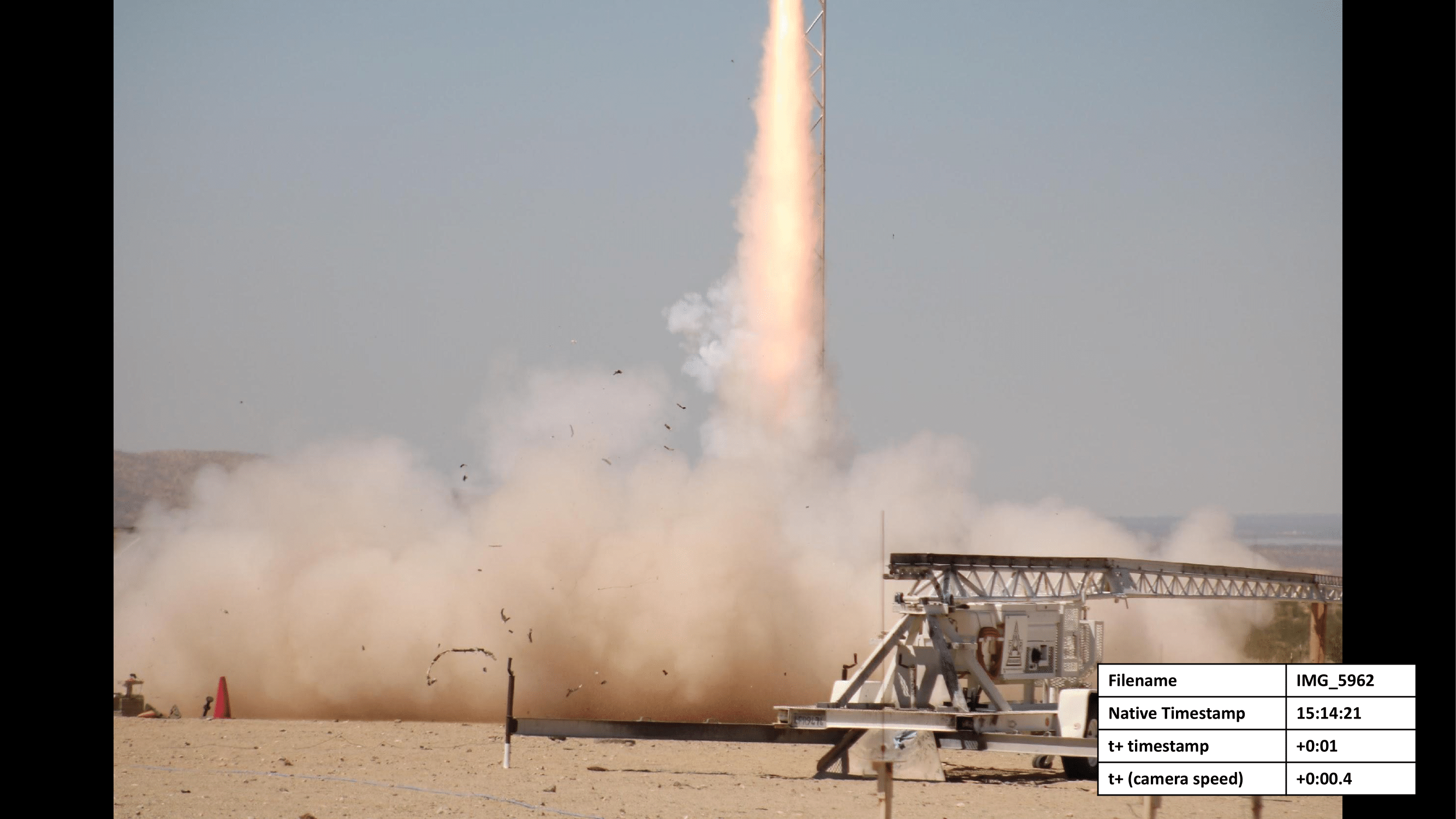}
\end{figure}

\begin{figure}[h!]
    \centering
    \includegraphics[width=0.9\textwidth]{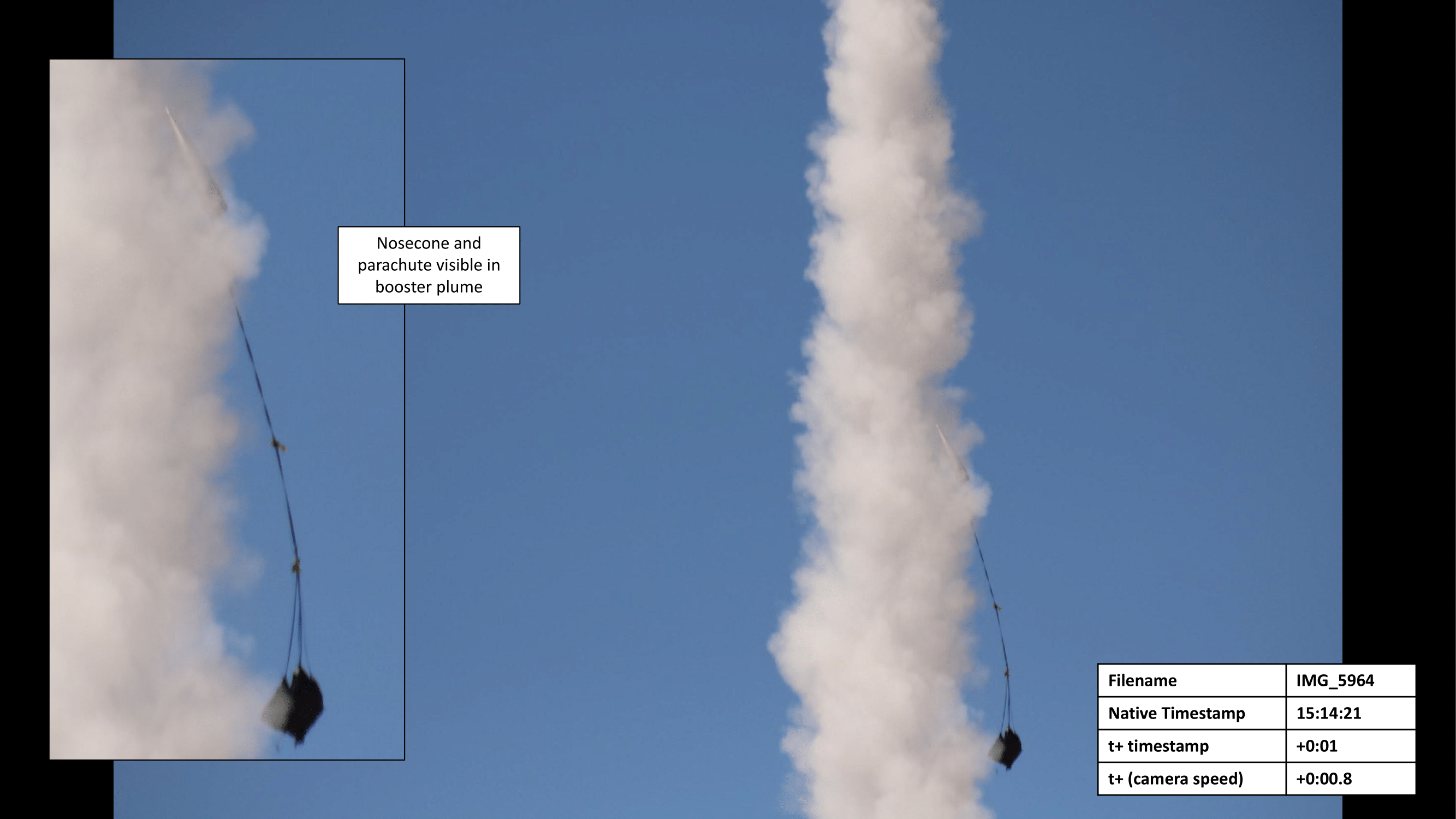}
\end{figure}
\begin{figure}[h!]
    \centering
    \includegraphics[width=0.9\textwidth]{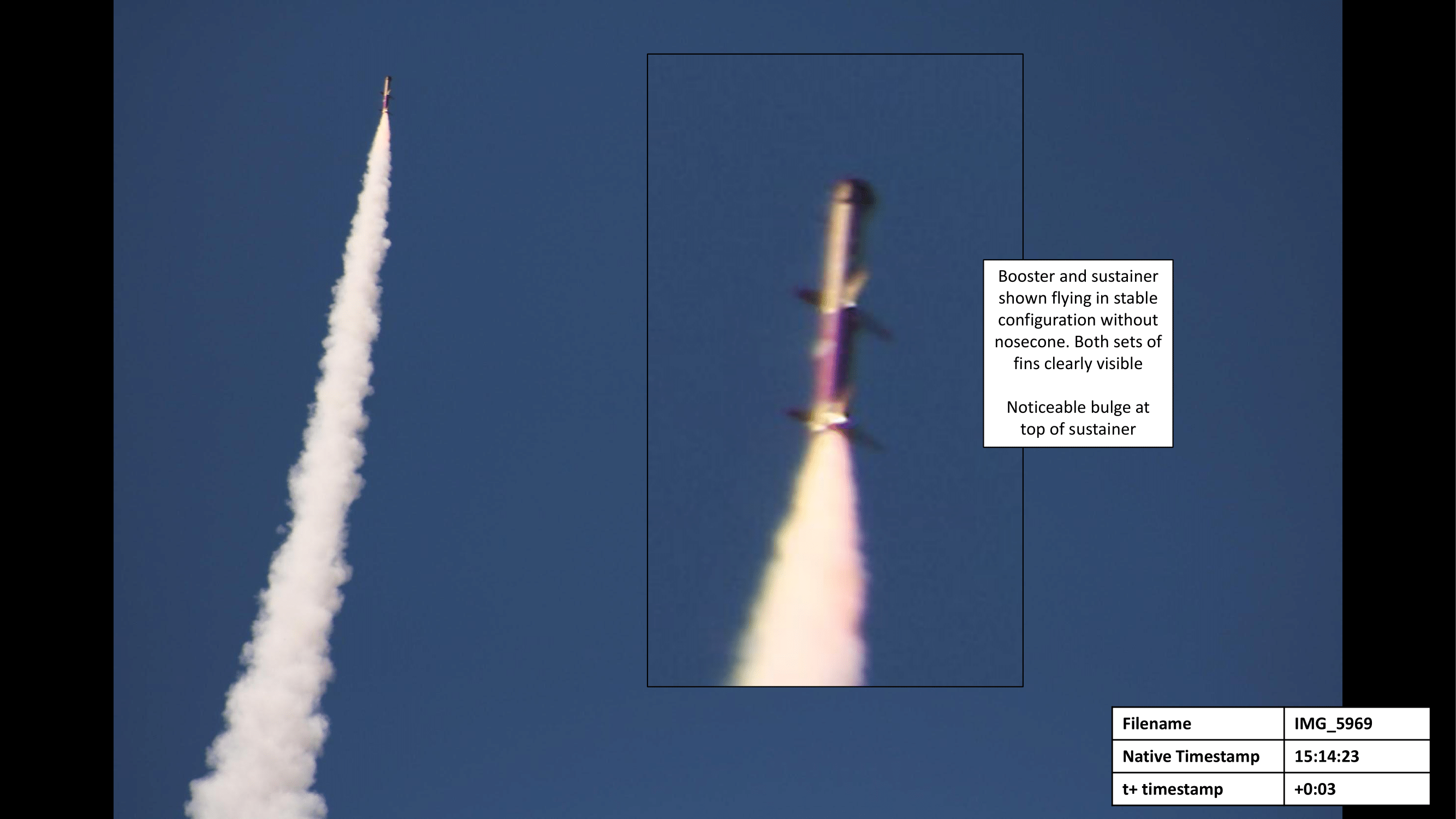}
\end{figure}

\begin{figure}[h!]
    \centering
    \includegraphics[width=0.9\textwidth]{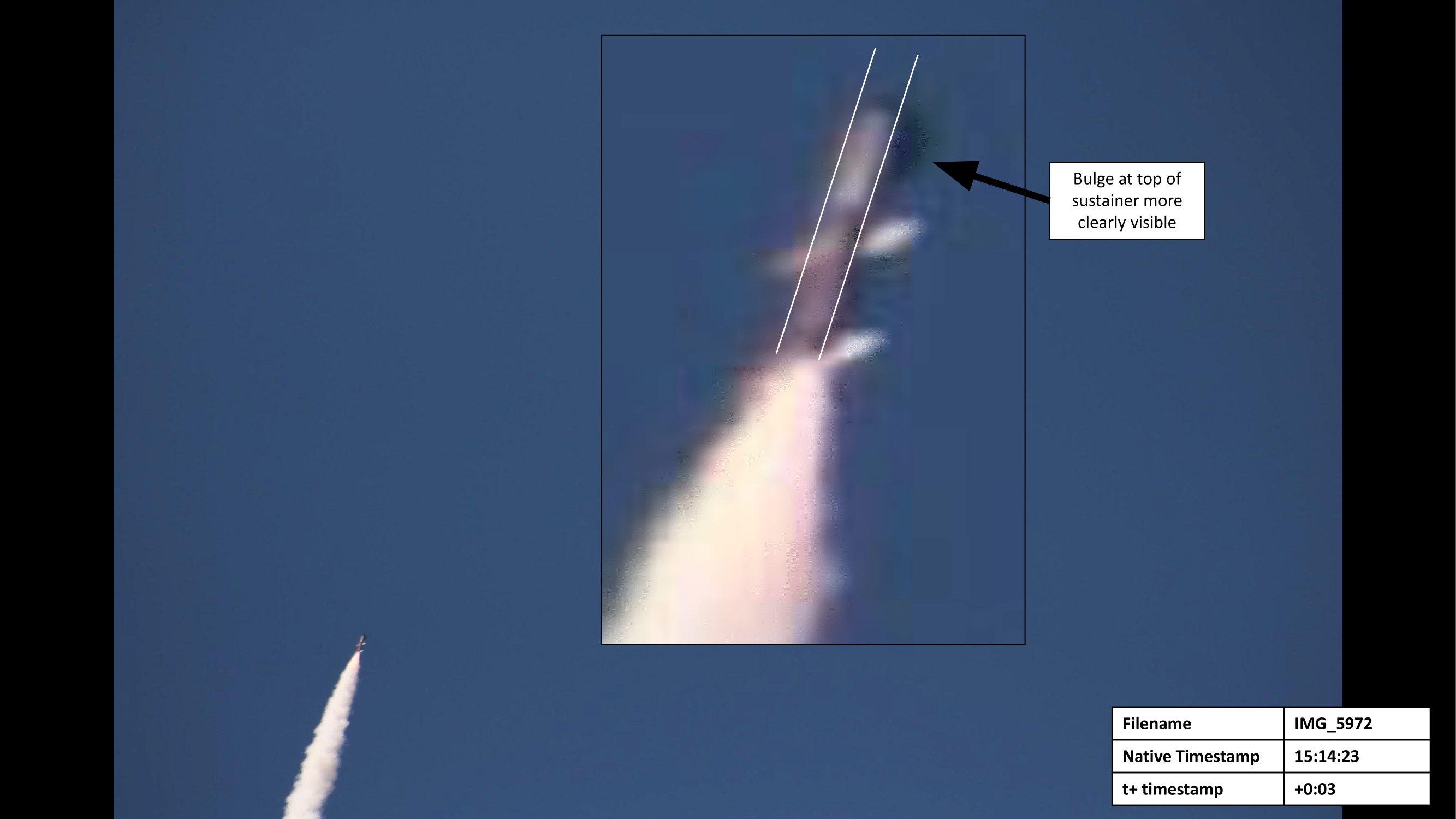}
\end{figure}

\begin{figure}[h!]
    \centering
    \includegraphics[width=0.9\textwidth]{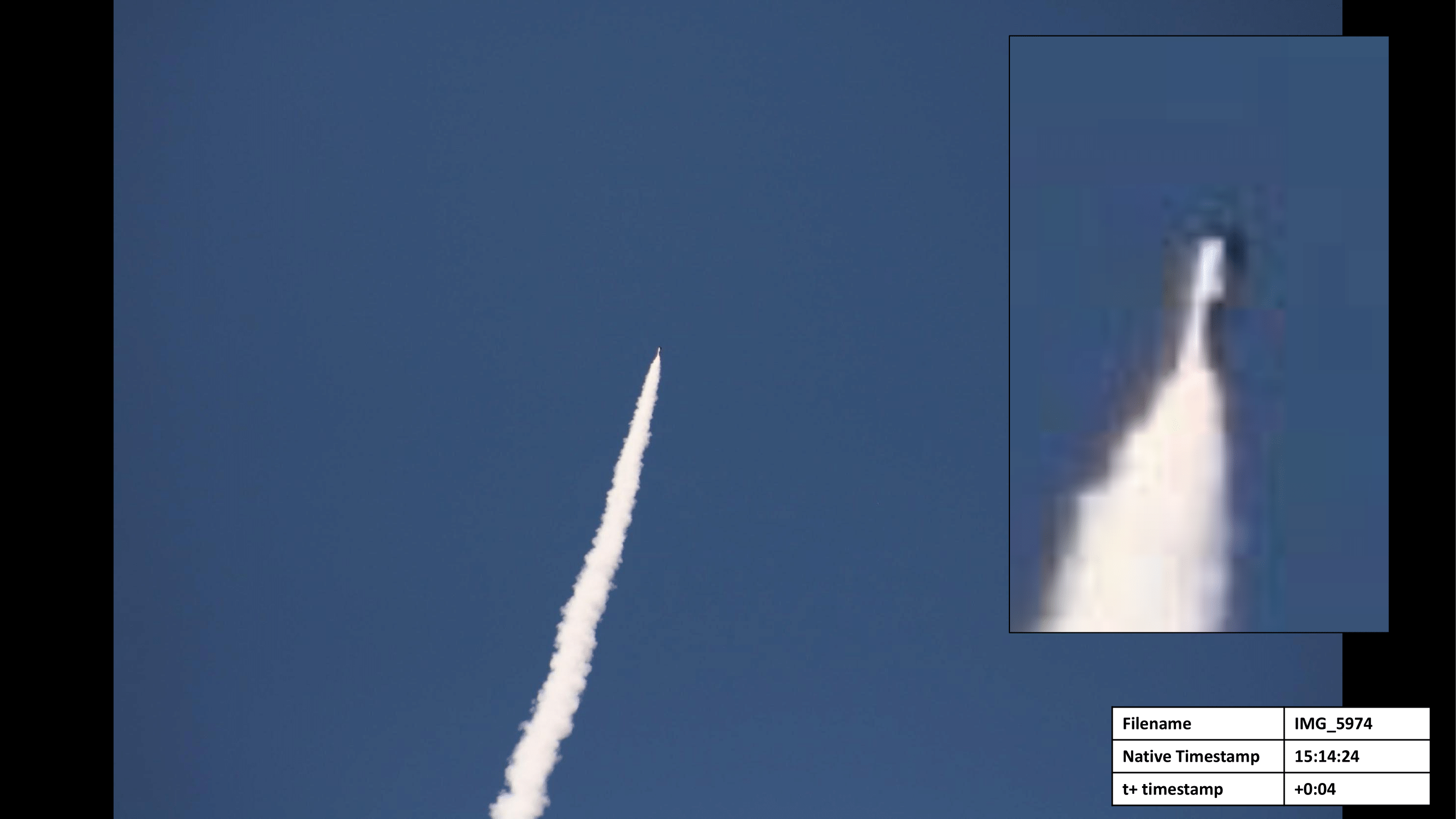}
\end{figure}
\begin{figure}[h!]
    \centering
    \includegraphics[width=0.9\textwidth]{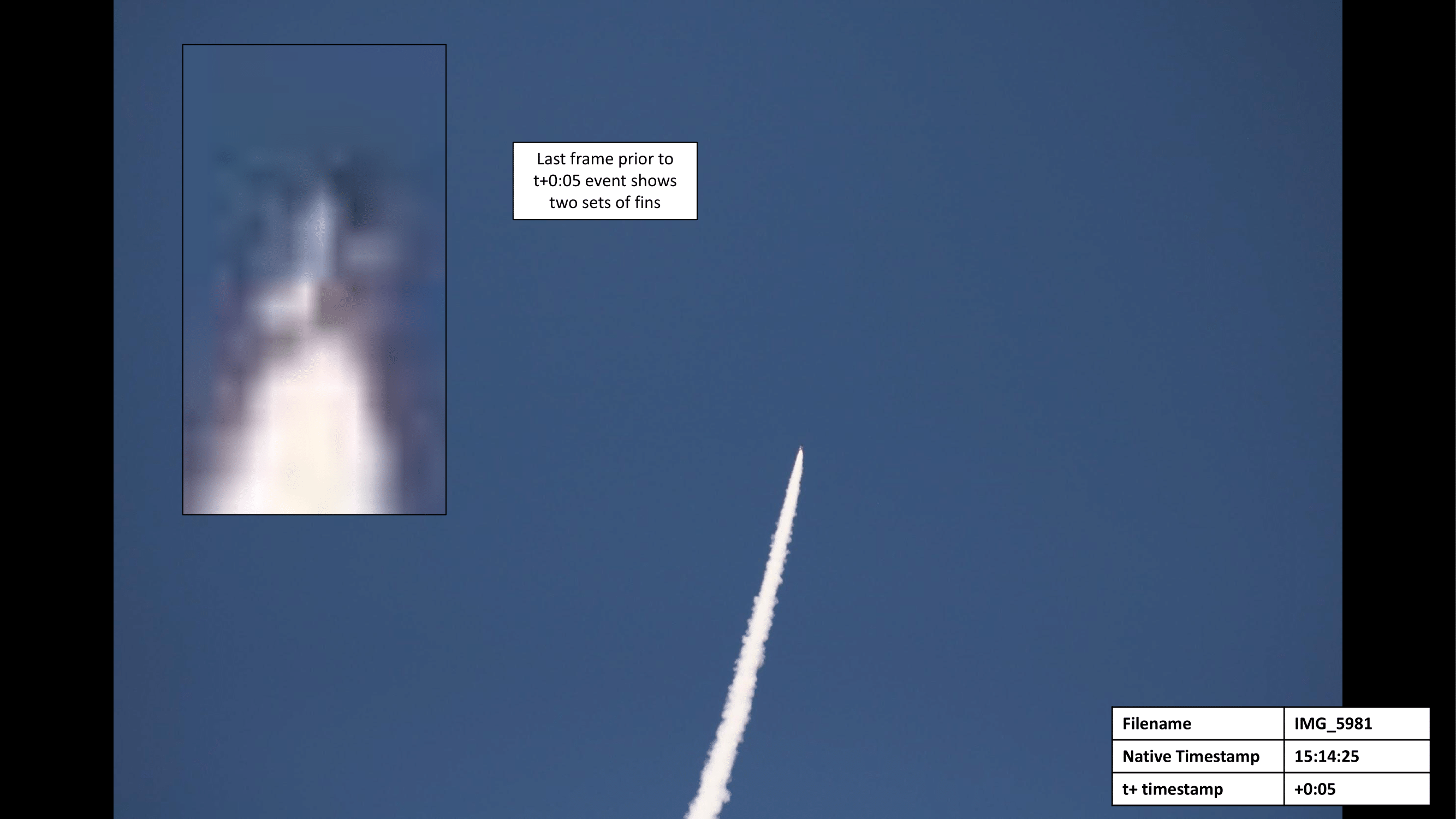}
\end{figure}

\begin{figure}[h!]
    \centering
    \includegraphics[width=0.9\textwidth]{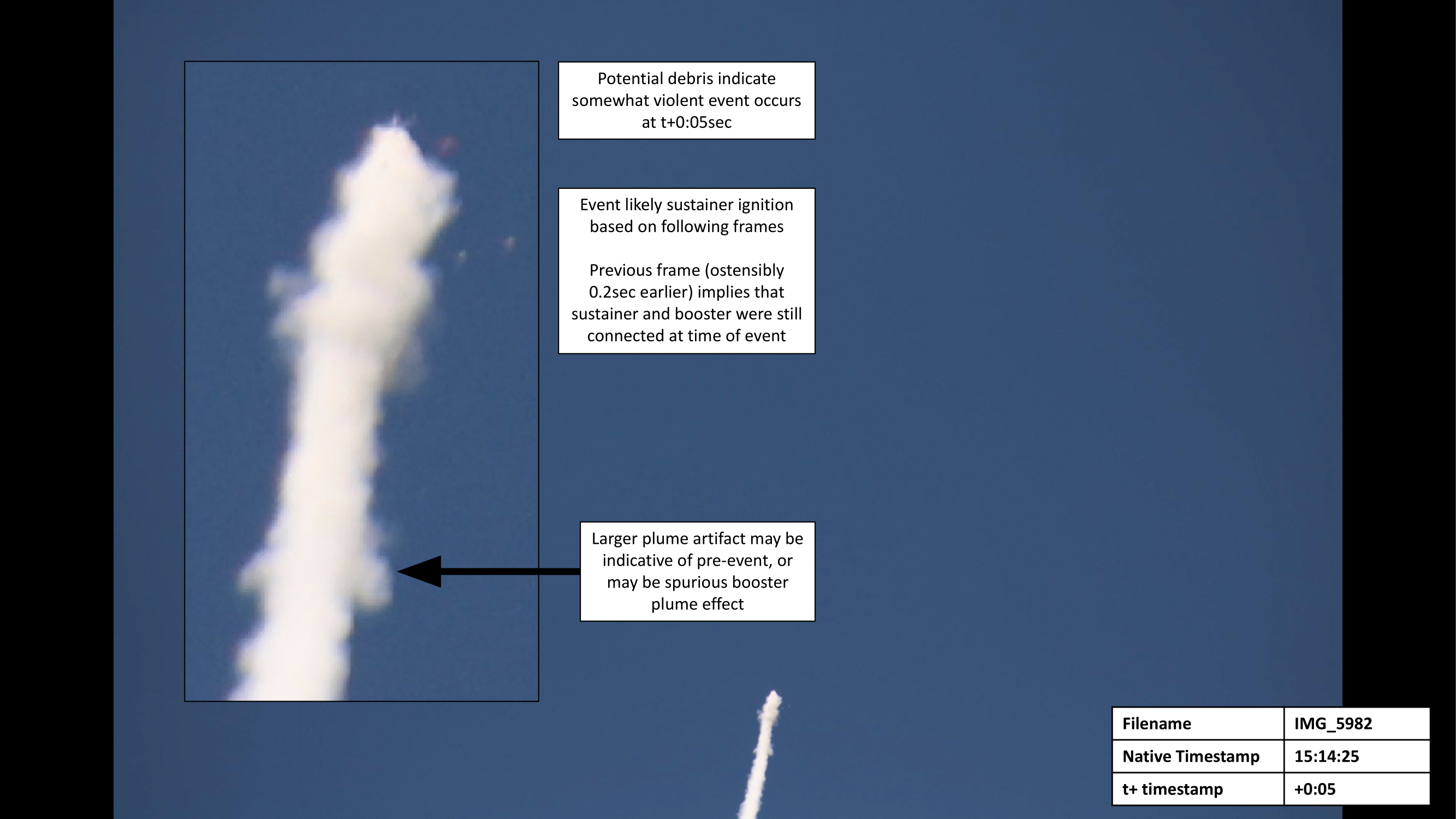}
\end{figure}

\begin{figure}[h!]
    \centering
    \includegraphics[width=0.9\textwidth]{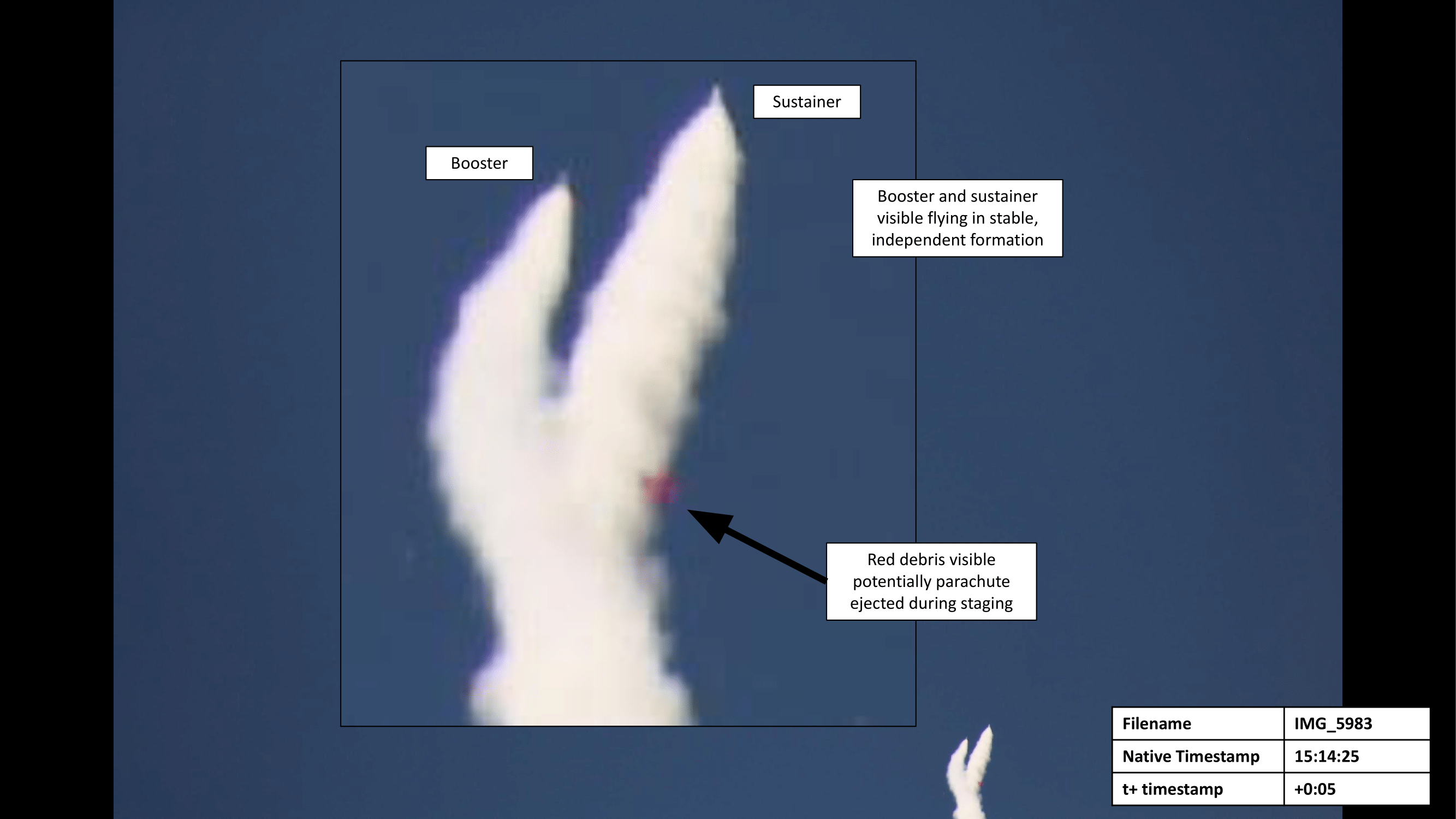}
\end{figure}

\begin{figure}[h!]
    \centering
    \includegraphics[width=0.9\textwidth]{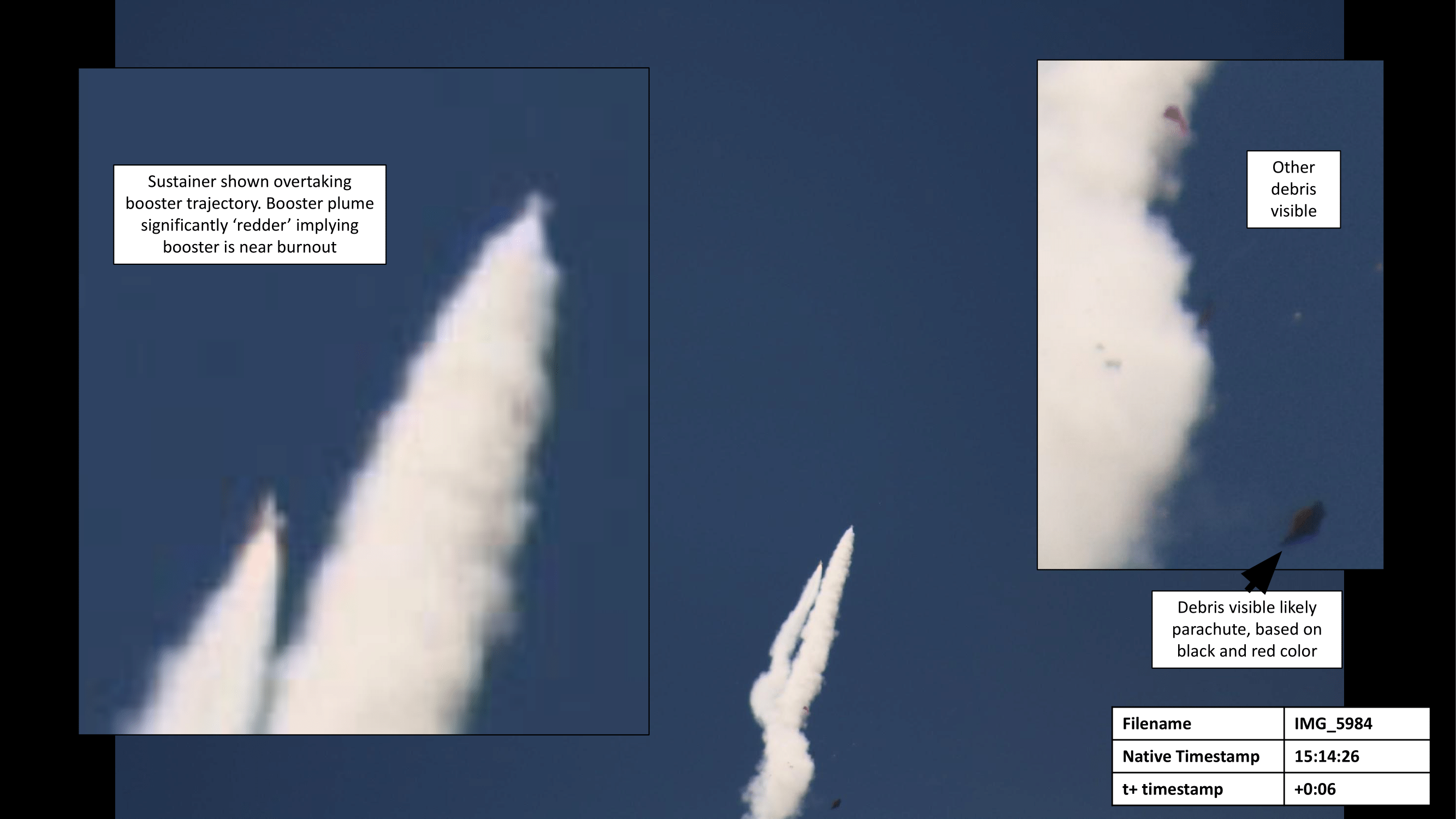}
\end{figure}
\begin{figure}[h!]
    \centering
    \includegraphics[width=0.9\textwidth]{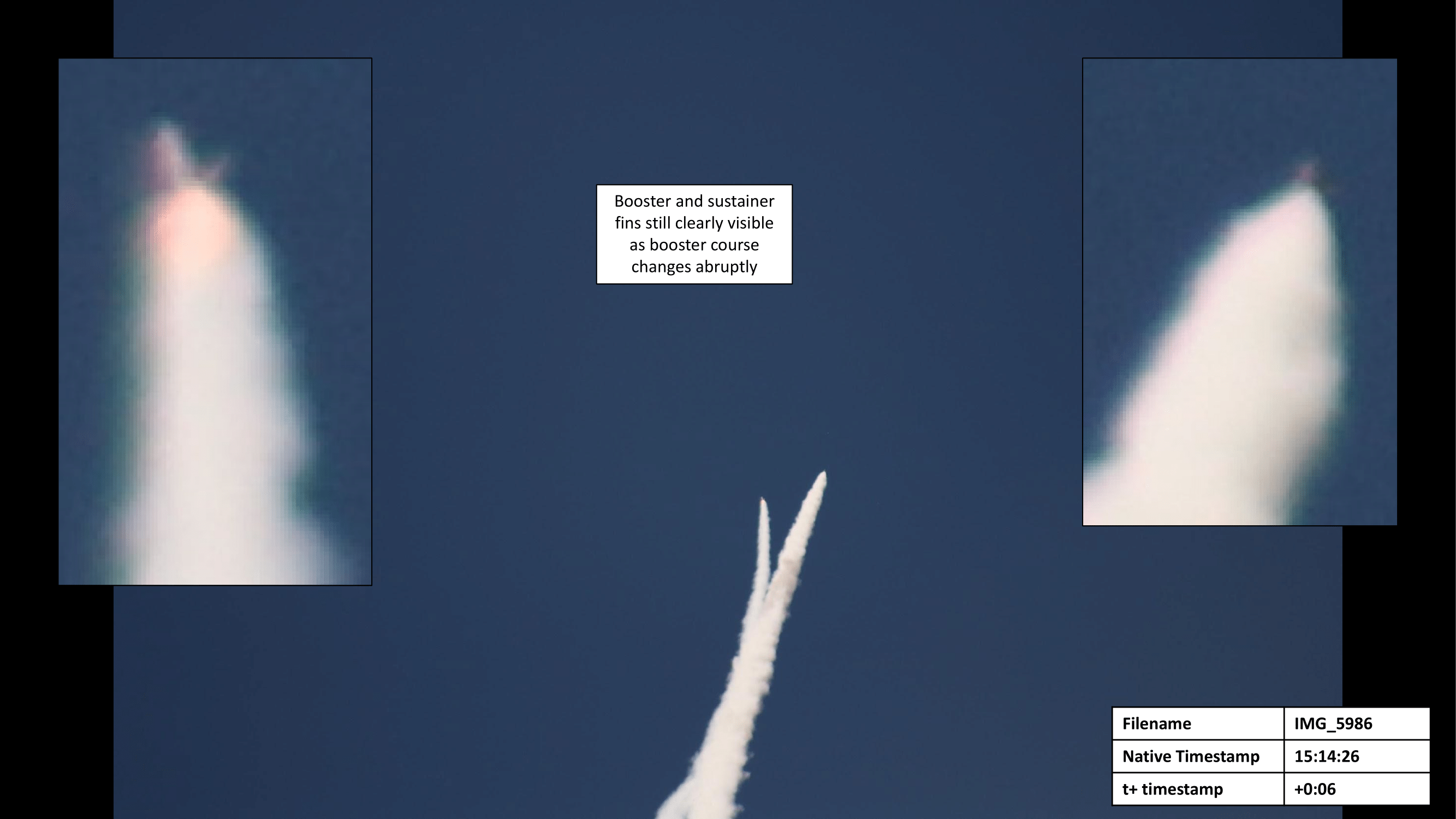}
\end{figure}

\begin{figure}[h!]
    \centering
    \includegraphics[width=0.9\textwidth]{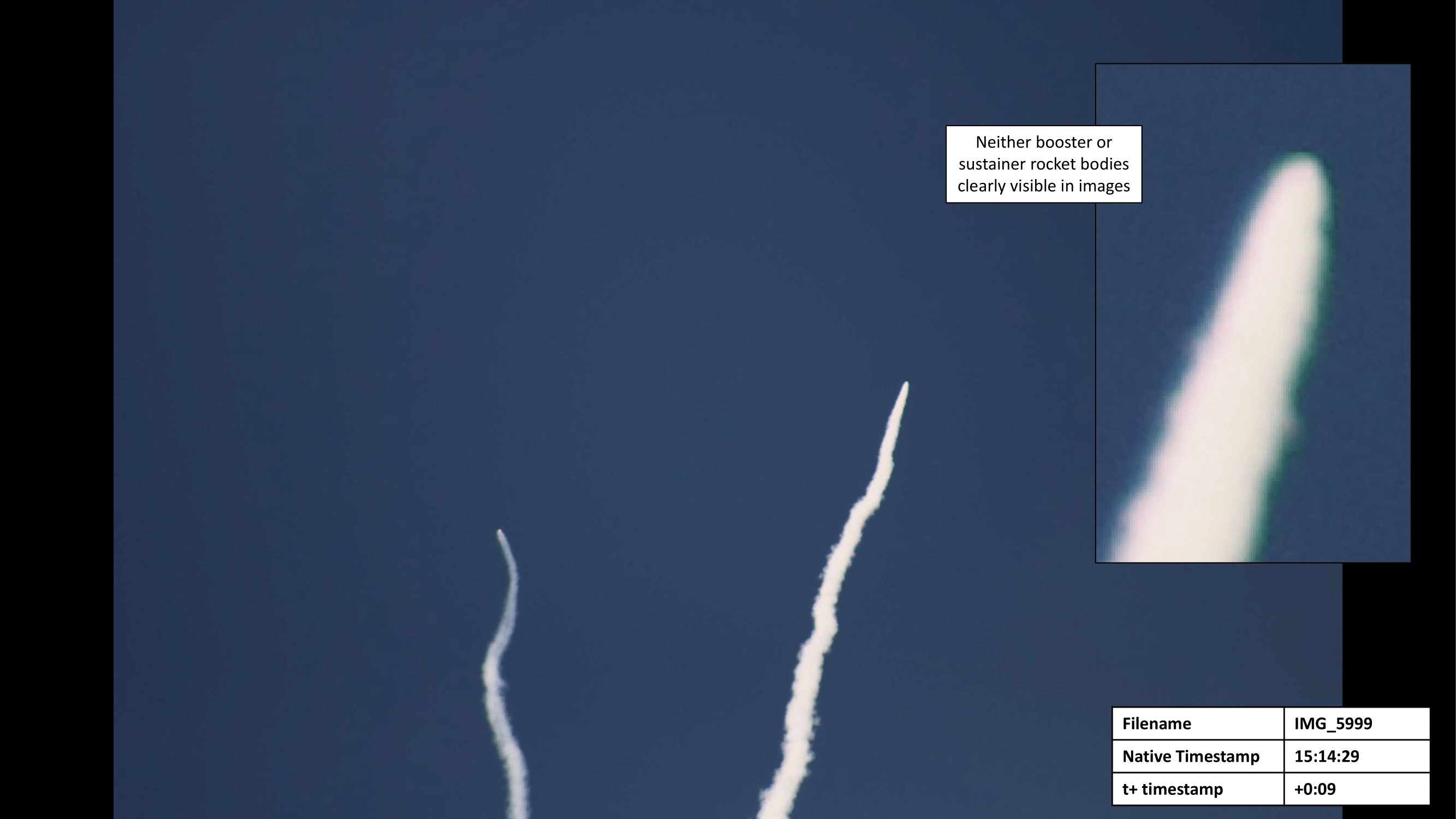}
\end{figure}

\begin{figure}[h!]
    \centering
    \includegraphics[width=0.9\textwidth]{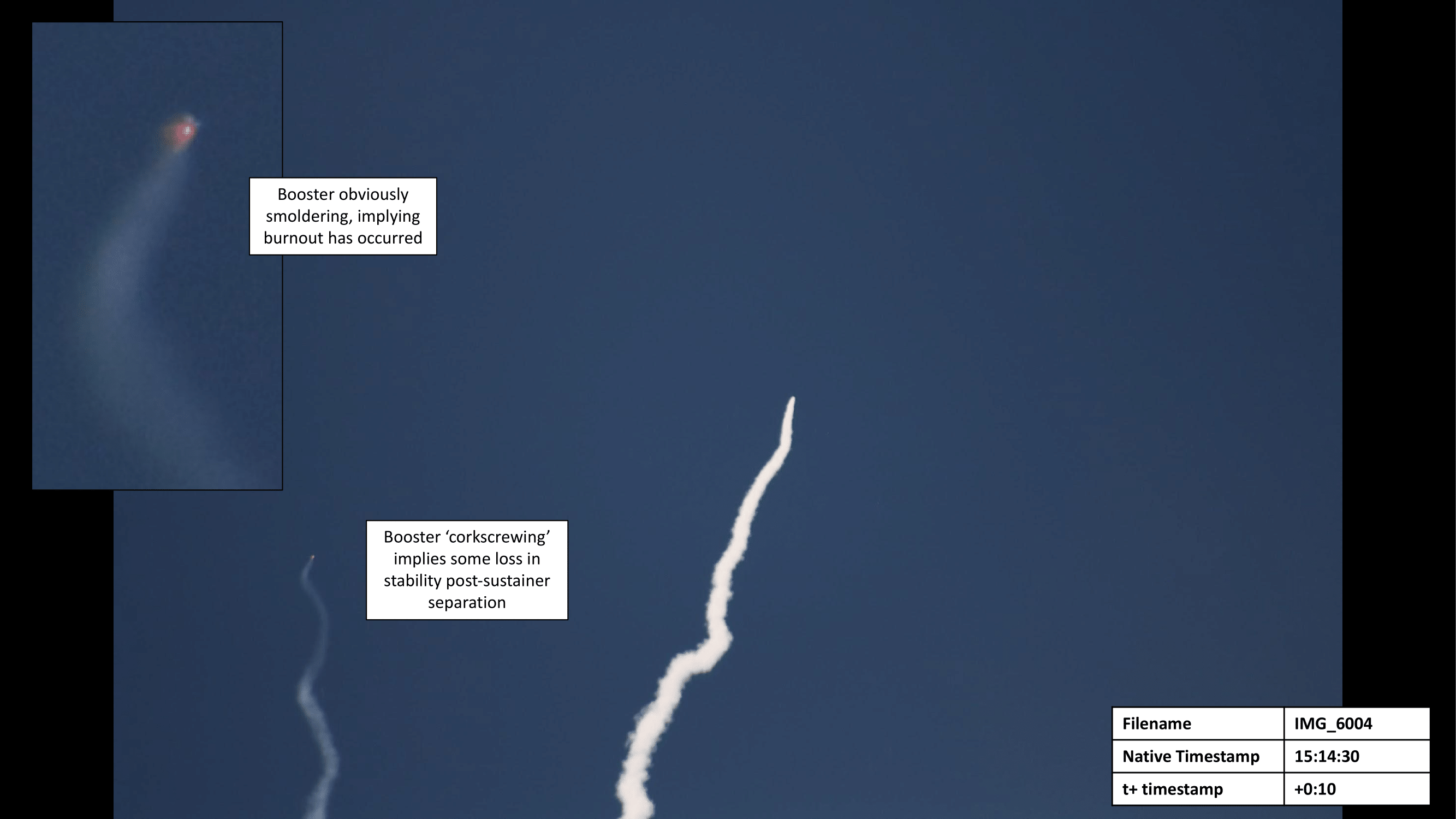}
\end{figure}

\begin{figure}[h!]
    \centering
    \includegraphics[width=0.9\textwidth]{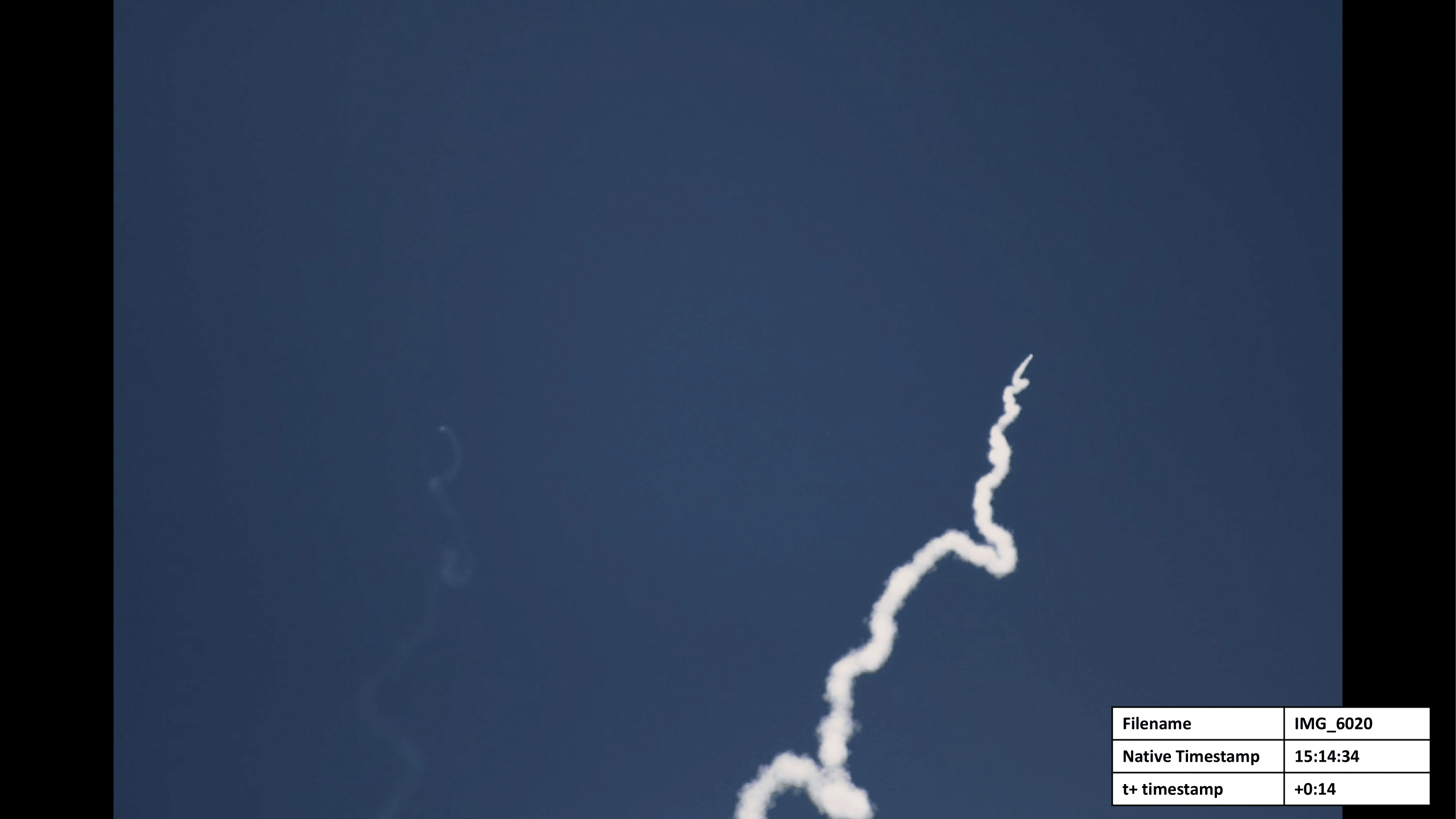}
\end{figure}

\begin{figure}[h!]
    \centering
    \includegraphics[width=0.9\textwidth]{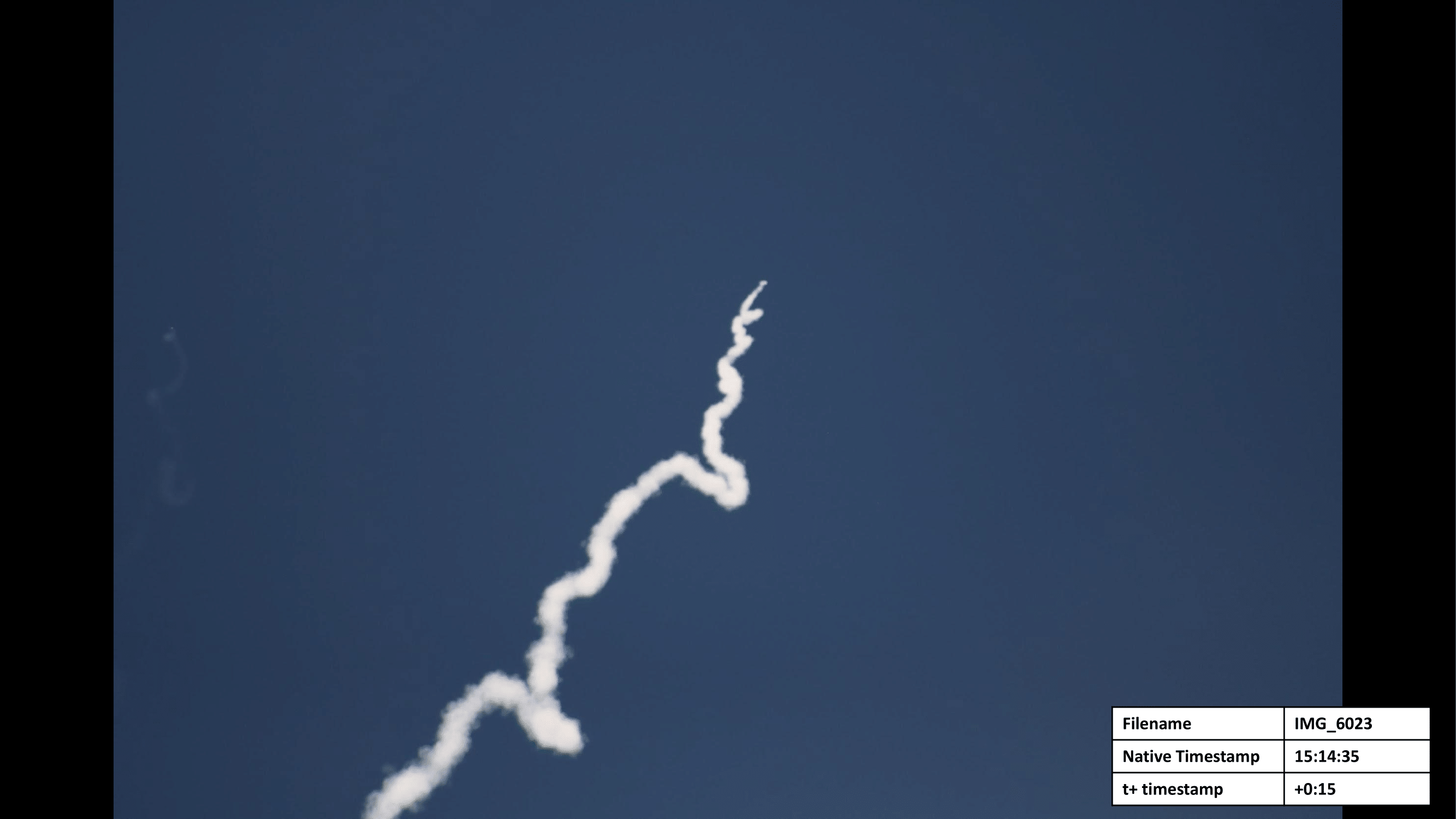}
\end{figure}
\begin{figure}[h!]
    \centering
    \includegraphics[width=0.9\textwidth]{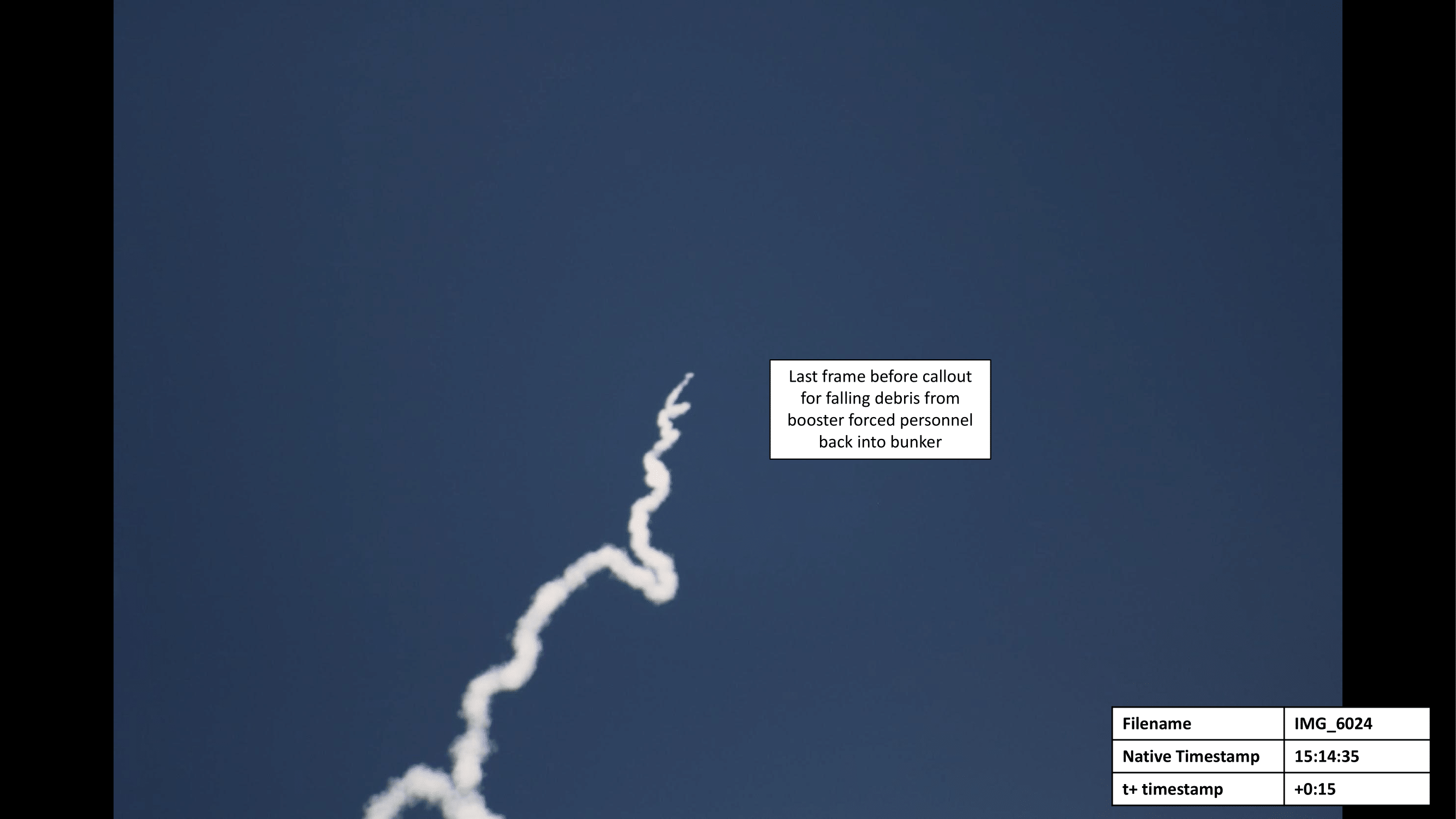}
\end{figure}

\begin{figure}[h!]
    \centering
    \includegraphics[width=0.9\textwidth]{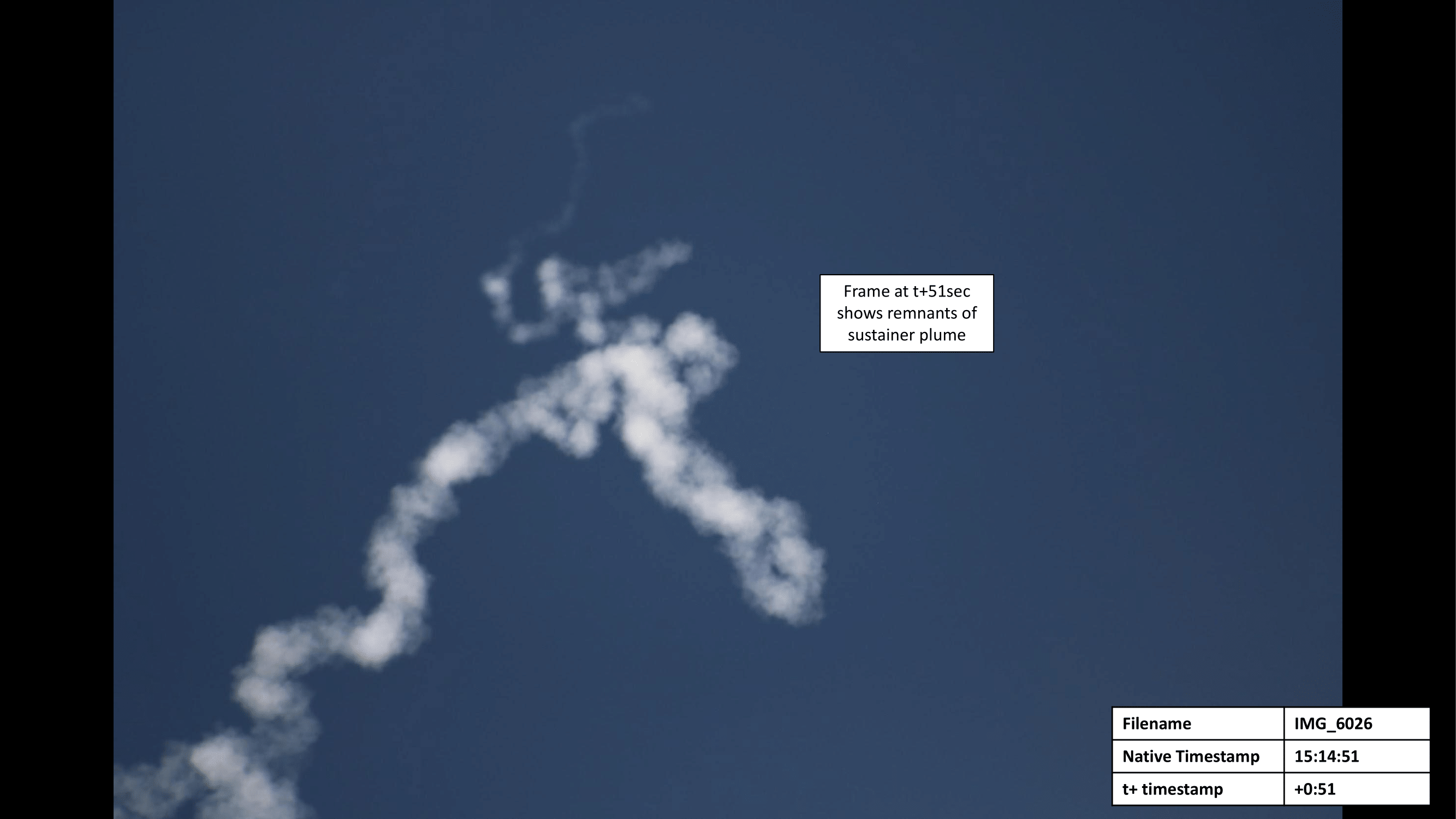}
\end{figure}

\begin{figure}[h!]
    \centering
    \includegraphics[width=0.9\textwidth]{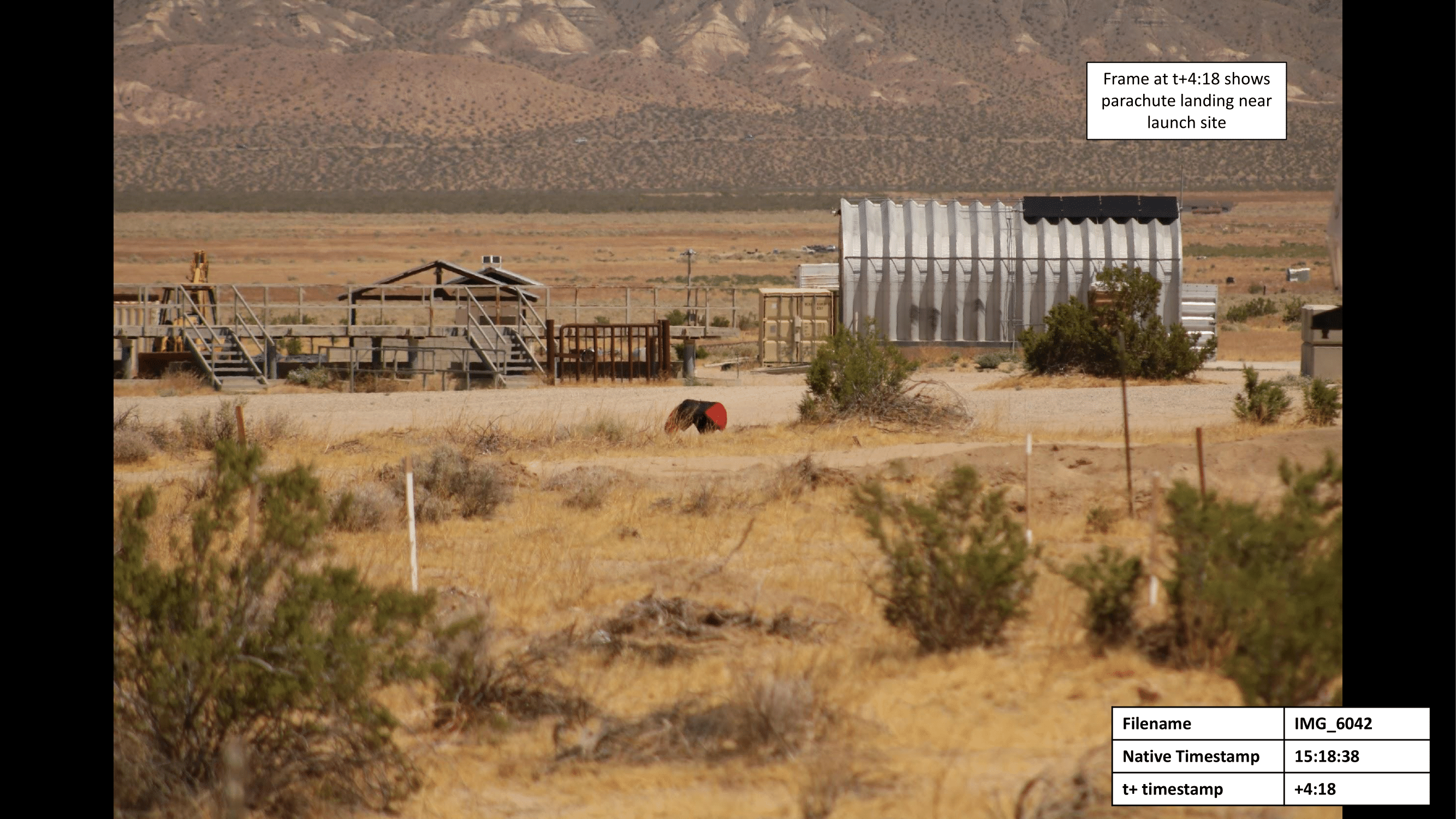}
\end{figure}

\end{document}